\documentclass[structabstract]{aa}  
\usepackage{balance}
\usepackage{wrapfig}
\usepackage{latexsym}                                   
\usepackage{longtable}
\usepackage{graphicx}
\usepackage{natbib}
\usepackage{natbib}
\bibpunct{(}{)}{;}{a}{}{,}
\usepackage{txfonts}
\defcitealias{stonkute12}{Paper~I}
\defcitealias{stonkute13}{Paper~II}
\begin{document}
   \title{Stellar substructures in the solar neighbourhood}

\subtitle{III. Kinematic group 2 in the Geneva-Copenhagen survey}

   \author{R. \v{Z}enovien\.{e}\inst{1},
          G. Tautvai\v{s}ien\.{e}\inst{1},
          B. Nordstr\"{o}m\inst{2},
          \and
          E. Stonkut\.{e}\inst{1}
          }
   \institute{Institute of Theoretical Physics and Astronomy (ITPA), Vilnius University,
              A. Gostauto 12, LT-01108 Vilnius, Lithuania\\
              \email{[renata.zenoviene;grazina.tautvaisiene;edita.stonkute]@tfai.vu.lt}
         \and
        Niels Bohr Institute, Copenhagen University, Juliane Maries Vej 30, DK-2100, Copenhagen, Denmark\\
             \email{birgitta@nbi.ku.dk}
             }

  \date{Received December XX, 2013; accepted December XX, 2013}

 
  \abstract
   {From correlations between orbital 
parameters, several new coherent groups of stars were recently identified in the Galactic disc and suggested to correspond to 
remnants of disrupted satellites. To reconstruct their origin at least three main observational parameters -- kinematics, chemical 
composition and age -- must be known. }
   {We determine detailed elemental abundances in stars belonging to the so-called Group~2 of the Geneva-Copenhagen Survey and 
   compare the chemical composition with Galactic thin- and thick-disc stars, as well as with the Arcturus and AF06 streams (\citealt{arifyanto06}). The aim is to search for 
   chemical signatures that might give information 
   about the formation history of this kinematic group of stars.}
   {High-resolution spectra were obtained with the FIES spectrograph at the Nordic Optical Telescope, La Palma, and were analysed
with a differential model atmosphere method. Comparison stars were observed and analysed with the same method.}
   {The average value of [Fe/H] for the 32 stars of Group~2 is $-0.42\pm 0.10$~dex. The investigated group consists mainly of two 8- and 
   12-Gyr-old stellar populations.  Abundances of oxygen, $\alpha$-elements, and r-process-dominated elements are 
   higher than in Galactic thin-disc dwarfs. This elemental abundance pattern 
   has similar characteristics as that of the  Galactic thick-disc.}
{The similarity in chemical composition of stars in Group~2 with that in stars of the thick-disc might suggest that their 
formation histories are linked. The chemical composition together with the kinematic properties and ages of stars in the 
investigated stars provides evidence of their common origin and possible relation 
to an ancient merging event.  A gas-rich satellite merger scenario is proposed as the most likely origin. Groups 2 and 3 of the 
Geneva-Copenhagen Survey might have originated in the same merging event.  
}


\titlerunning{Stellar substructures in the solar neighbourhood.III.}
\authorrunning{R. \v{Z}enovien\.{e} et al. }

\maketitle


\section{Introduction}

A combined study of kinematics and chemical composition of stars is one 
of the most promising tools of research in galaxy formation. The main goal in this field of research is to reconstruct the formation history 
of our Galaxy, to reveal the origin of the thick disc, and to find remnants of ancient mergers. Large surveys, such as the Geneva Copenhagen Survey (GCS, 
\citealt{nordstrom04}), the RAdial Velocity Experiment (RAVE, \citealt{steinmetz06, zwitter08, siebert11}), and the Sloan 
Extension for Galactic Understanding and Exploration (SEGUE, \citealt{yanny09}), are very good databases for such studies. 

   \begin{figure*}
   \centering
   \includegraphics[width=\textwidth]{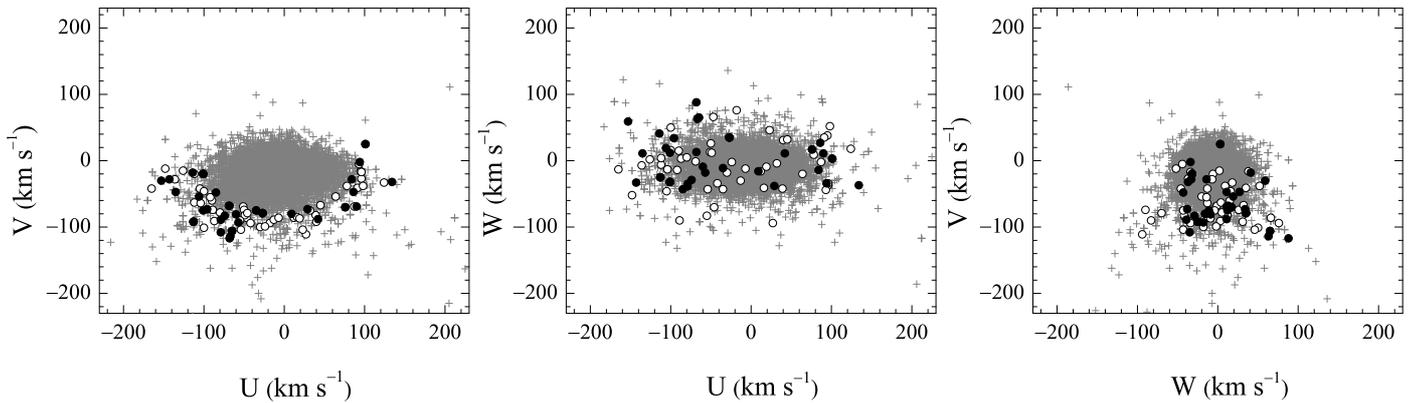}
      \caption{Velocity distribution for all stars in the sample of \citet{holmberg09} (plus signs), stars of Group~2 (circles), and the investigated stars (filled circles). 
              }
         \label{Fig.1}
   \end{figure*}

   \begin{figure*}
   \centering
 \includegraphics[width=0.95\textwidth]{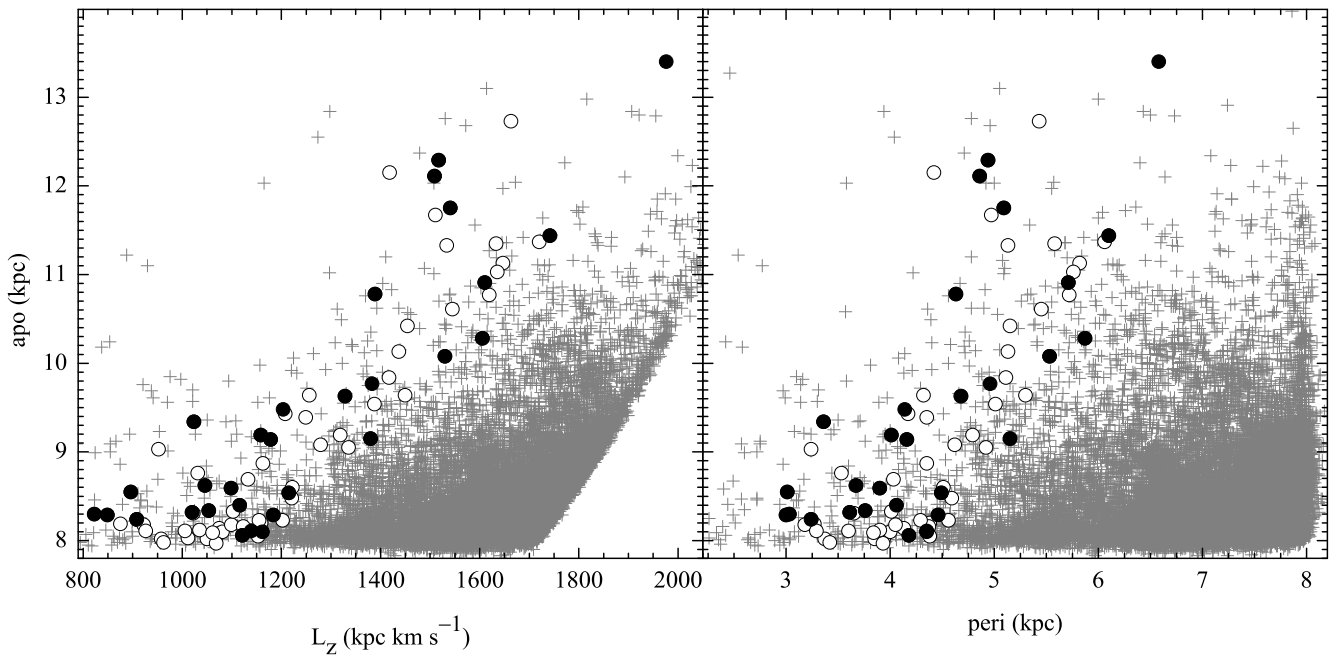}

      \caption{Distribution for the stars in the APL space. Plus signs denote the GCS sample \citep{holmberg09}, circles denote Group~2, the filled 
      circles are the investigated stars. Note that the investigated stars as well as all Group 2 stars are distributed
      in APL space with constant eccentricity. 
              }
         \label{Fig.2}
   \end{figure*}

In this series of papers, we focus on the detailed chemical composition study of recently identified  stellar kinematic groups that were suggested 
to belong to remnants of ancient merger events in our Galaxy.  The GCS Group~3 was investigated recently by \citet[hereafter Paper~I and Paper~II] 
{stonkute12, stonkute13}, while here we present our results of analysis of the GCS Group~2. These kinematic 
stellar groups were identified in the GCS catalogue by \citet{helmi06}.   These authors analysed numerical simulations of the disruption of satellite galaxies 
and found substructures in the space defined by apocentre, 
pericentre, and $z$-angular momentum (the so called APL-space).  Stars released in different perigalactic passages of the merging galaxy have slightly 
different orbital properties, and hence are located in several smaller lumps in the APL-space. However, these lumps are located 
along a segment of constant eccentricity, thereby permitting the assessment of a common origin. Such substructures in the APL-space remain coherent 
for many Gyr, well after the mergers have fully mixed. The APL-space for the GCS catalogue shows large amounts of substructure. The most prominent 
structures are related to the superclusters Hyades-Pleiades, Sirius and Hercules, and are most likely generated by dynamical perturbations induced by the 
spiral arms and the Galactic bar. These structures are composed of stars on disc-like orbits with relatively low eccentricity. However, the detailed 
statistical analysis of the APL-space revealed about ten other overdensities at significance levels higher than 99\%. These 
overdensities were located along two to three segments of constant eccentricity, as predicted for substructures that are the result of minor mergers. 
There were 274 stars in this region of the APL-space, which is delimited by eccentricity $ 0.3 \leq \epsilon < 0.5$. 
The metallicity distribution of the stars in this overdense region of the APL-space varied with eccentricity in a discontinuous fashion. This allowed the 
separation of these stars into three Groups. These three groups of stars are dissimilar not only in their metallicity distribution, but also have different 
kinematics in the vertical ($z$) direction.   The Group~1 velocity dispersion has $\sigma_z$  about $28\,{\rm km\,s}^{-1}$, that of Group~2 about 
$39\,{\rm km\,s}^{-1}$, and that of Group~3 about $52\,{\rm km\,s}^{-1}$. The groups also have distinct age distributions. Group~2 contains 86 stars; for 32 
of them we determine the detailed chemical composition in this work. 

\section{Observations and analysis}

 A list of the observed stars and some of their parameters (taken from the catalogue of 
\citealt{holmberg09} and Simbad) are presented in Table~\ref{table:1}. Thirty-two stars of Group~2 and seven comparison stars    
(thin-disc dwarfs) were observed.

\begin{figure}
   \centering
   \includegraphics[width=0.4\textwidth]{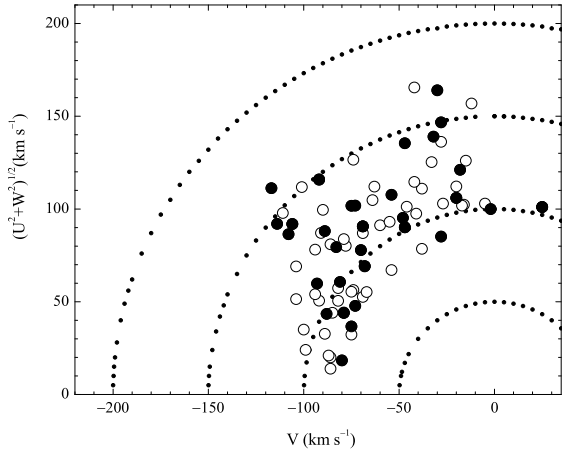}
\caption{Toomre diagram of all stars of Group 2 (circles) and 
those investigated in this work (filled circles). Dotted lines 
indicate constant values of total space velocity in steps of 50 km s$^{-1}$.} 
\label{Fig.3}
\end{figure}

In Fig.~\ref{Fig.1} we show the Galactic disc stars from \citet{holmberg09}. Stars belonging to Group~2 in \citet{helmi06} 
are marked with open and filled circles (the latter are used to mark stars investigated in our work). The stars belonging to Group 2 
have a different distribution in the velocity space than other stars of the Galactic disc. 
In Fig.~\ref{Fig.2} the stars are shown in the APL-space. The Toomre diagram of all stars of Group 2 and 
those investigated in this work are presented in Fig.~\ref{Fig.3}.  

\begin{figure}
   \centering
   \includegraphics[width=0.47\textwidth]{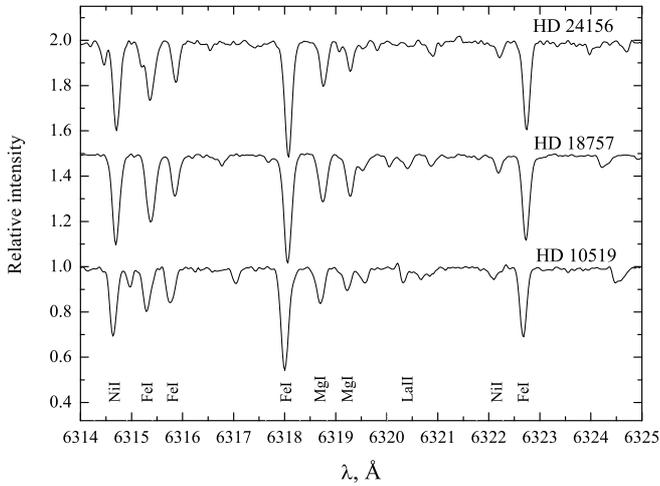}
\caption{Samples of stellar spectra of several programme stars. An offset of 0.5 in relative flux is applied for clarity.} 
\label{Fig.4}
\end{figure}

Spectra of high-resolving power ($R\approx68\,000$) in the wavelength range of 3680--7270~{\AA} were obtained 
at the Nordic Optical Telescope with the FIES  spectrograph during July 2008. All spectra were exposed to reach a signal-to-noise 
ratio higher than 100. Reductions of CCD images were made 
with the FIES pipeline FIEStool, which performs  a complete reduction: calculation of reference frame, bias and 
scattering subtraction, flat-field dividing, wavelength calibration and other procedures (http://www.not.iac.es/instruments/fies/fiestool). 
Several examples of observed stellar spectra are presented in Fig.~\ref{Fig.4}.

The spectra were analysed using the differential model atmosphere technique described in Papers~I and II. 
Here we only recall some details. The Eqwidth and BSYN program packages, developed at the Uppsala Astronomical 
Observatory, were used to carry out the calculation of abundances from measured 
equivalent widths and synthetic spectra, respectively.
A set of plane-parallel, line-blanketed, constant-flux LTE model atmospheres 
\citep{gustafsson08} were taken from the MARCS stellar model atmosphere and flux 
library (http://marcs.astro.uu.se/).

\begin{table*}
	\centering
\begin{minipage}{150mm}
\caption{Parameters of the programme and comparison stars}
\label{table:1}
\begin{tabular}{lcrrrrrcrrr}
\hline  
\hline
         Star &  Sp. type &	 $M_{\rm V}$ & $d$ & $U$ &  $V$ &  $W$ & $e$ & $z_{\rm max}$ & $R_{\min}$ & $R_{\max}$  \\
   			&	 &   mag           & pc & km s$^{-1}$   &  km s$^{-1}$ & km s$^{-1}$  &        & kpc & kpc & kpc \\
\hline
\noalign{\smallskip}
\object{BD +68 813}  & G5  & 4.27 &	126	&	--143	&	--28	&	--33	&	0.40 &	0.56	&	5.09	&	11.75	\\ 
\object{BD +31 3330} & K3V & 6.66 &	24	&	94	    &	--2		&	--34	&	0.30 &	0.55	&	6.10    &	11.44	\\ 
\object{HD 10519} & G2/G3V & 4.07 &	48	&	--96	&	--73	&	34      &   0.37 &	0.89	&	4.16	&	9.14	\\ 
\object{HD 12782}$^{*}$ &	G5 & 5.54 & 	37  &	 76    	&	--70    &	17      &	0.37 &	0.45	&	4.23    &	9.11   	\\ 
\object{HD 16397}  	& G0V  & 4.68 &	34	&	134		&	--32	&	--37	&	0.43 &	0.69	&	4.86	&	12.11	\\ 
\object{HD 18757}  	& G4V  & 4.71 &	24	&	--74	&	--83	&	--29	&	0.38 &	0.37	&	3.90	&	8.59	\\ 
\object{HD 21543} &	G2V-VI & 4.88 &	47	&	--57	&	--93	&	--18	&	0.39 &	0.17	&	3.61	&	8.32	\\ 
\object{HD 24156} &	G0	   & 3.82 &	74	&	29		&	--73	&	--38	&	0.30 &	0.56	&	4.46	&	8.29	\\ 
\object{HD 29587} &	G2V	   & 5.06 &	28	&	--135	&	--47	&	11		&	0.40 &	0.35	&	4.63	&	10.78	\\ 
\object{HD 30649} &	G1V-VI & 4.56 &	30	&	--60	&	--81	&	--9		&	0.35 &	0.03	&	4.06	&	8.40	\\ 
\object{HD 37739} &	F5 	   & 3.48 &	77	&	--27	&	--79	&	35	    &	0.30 &	0.84	&	4.35	&	8.11	\\
\object{HD 38767} &	F8     & 3.35 &	71	&	101		&	25		&	3		&	0.34 &	0.19	&	6.58	&	13.40	\\ 
\object{HD 96094} &	G0     & 3.90 &	55	&	--85	&	--48	&	--43	&	0.28 &	0.71	&	5.15	&	9.15	\\ 
\object{HD 114606} & G1V &	4.87 &	59	&	--153	&	--30	&	59		&	0.43 &	2.17	&	4.94	&	12.29	\\ 
\object{HD 121533} & G5	&	4.71 &	59	&	42		&	--88	&	11		&	0.38 &	0.31	&	3.76	&	8.34	\\ 
\object{HD 131582} & K3V &	6.76 &	24	&	--68	&	--68	&	13		&	0.31 &	0.34	&	4.49	&	8.54	\\ 
\object{HD 132142} & K1V &	5.91 &	23	&	--106	&	--54	&	19		&	0.35 &	0.48	&	4.68	&	9.63	\\ 
\object{HD 133621} & G0	&	3.98 &	34	&	--35	&	--75	&	--11	&	0.30 &	0.06	&	4.35	&	8.10	\\ 
\object{HD 137687} & G9	&	3.68 &	54	&	--79	&	--108	&	--35	&	0.48 &	0.51	&	3.01	&	8.55	\\ 
\object{HD 139457} & F8V &	3.82 &	45	&	84		&	--28	&	--14	&	0.29 &	0.12	&	5.53	&	10.08	\\ 
\object{HD 143291} & K0V &	5.91 &	26	&	--101	&	--75	&	12		&	0.39 &	0.33	&	4.01	&	9.19	\\ 
\object{HD 152123} & G5	&	2.69 &	153	&	--101	&	--20	&	--32	&	0.27 &	0.49	&	5.87	&	10.28	\\
\object{HD 156802}$^{*}$ & G2V & 3.66 &	72  &	--58    &	--84    &	--61    &	0.34 &	1.24 	&	4.03    &	8.28     \\ 
\object{HD 158226} & G1V &	4.31 &	69	&	--65	&	--106	&	65		&	0.43 &	2.09	&	3.24	&	8.24	\\ 
\object{HD 165401} & G0V &	4.91 &	24	&	--79	&	--89	&	--39	&	0.40 &	0.60	&	3.67	&	8.62	\\ 
\object{HD 170357} & G1V &	3.99 &	72	&	--68	&	--117	&	88		&	0.46 &	3.39	&	3.03	&	8.30	\\ 
\object{HD 190404} & K1V &	6.29 &	16	&	86		&	--47	&	27		&	0.33 &	0.73	&	4.96	&	9.77	\\ 
\object{HD 200580} & F9V &	3.74 &	52	&	90		&	--69	&	11		&	0.39 &	0.31	&	4.14	&	9.48	\\ 
\object{HD 201099} & G0 &	4.21 &	47	&	--114	&	--18	&	41		&	0.31 &	1.18	&	5.71	&	10.91	\\ 
\object{HD 215594} & G5	&	3.99 &	88	&	9		&	--80	&	--16	&	0.32 &	0.13	&	4.18	&	8.06	\\ 
\object{HD 221830} & F9V &	4.26 &	33	&	--67	&	--114	&	63		&	0.47 &	2.04	&	3.00    &   8.29	\\ 
\object{HD 224817} & G2V &	4.09 &	73	&	--113	&	--92	&	--25	&	0.47 &	0.31	&	3.36	&	9.34	\\
\noalign{\smallskip}
 \hline
 \noalign{\smallskip}
\object{HD 41330}	& G0V	& 4.08 &	26	&	10	&	--25	&	--32	&	0.10	&	0.40	&	6.75	&	8.21	\\
\object{HD 43318}	& F6V 	& 2.77 &	37	&	50	&	3	    &	--36	&	0.18	&	0.57	&	7.01	&	10.13	\\
\object{HD 69897}	& F6V	& 3.83 &	18	&	--24 &	--39	&	7	    &	0.14	&	0.21	&	6.09	&	8.07	\\
\object{HD 108954}	& F9V	& 4.52 &	22	&	0	&	8	    &	--28	&	0.06	&	0.34	&	7.95	&	9.01	\\
\object{HD 153597}	& F8V 	& 3.97 &	15	&	2	&	--9	    &	--29	&	0.04	&	0.34	&	7.61	&	8.19	\\
\object{HD 157466}	& F8V	& 4.55 &	29	&	40	&	16	    &	3	    &	0.17	&	0.18	&	7.37	&	10.37	\\
\object{HD 176377}	& G0 	& 4.92 &	24	&	--40 &	--25	&	--5 	&	0.12	&	0.04	&	6.57	&	8.34	\\
 
\hline

\end{tabular}

\end{minipage}
\tablefoot{$^*$ Data for HD\,12782 and HD\,156802 were taken from \citet{holmberg07}.}
\end{table*}

\begin{figure}
   \centering
   \includegraphics[width=0.47\textwidth]{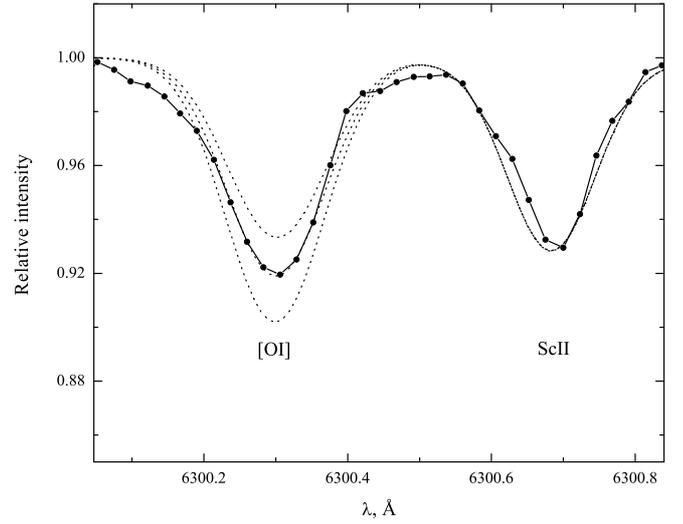}
\caption{Fit to the forbidden [O\,{\sc i}] line at 6300.3 {\AA} in the programme star HD 12782. 
The observed spectrum is shown as a solid line with black dots. The synthetic spectra with ${\rm [O/Fe]}=0.52 \pm 0.1$ 
are shown as dotted lines.} 
\label{Fig.5}
\end{figure}


   \begin{figure}
   \centering
   \includegraphics[width=0.47\textwidth]{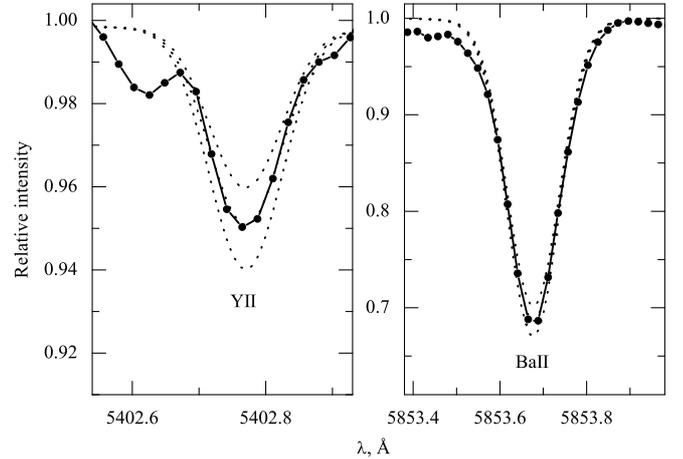}
      \caption{Synthetic spectrum fit to the yttrium line at 5402~{\AA} in the observed spectrum of HD 21543 (left panel) and to the barium line at 
      5853~{\AA} in the observed spectrum of HD 224817. The observed spectra are shown by solid lines with dots. The dotted lines are synthetic 
      spectra with ${\rm [Y/Fe]} = 0.01 \pm 0.1$ and with ${\rm [Ba/Fe]} = -0.10 \pm 0.1$, respectively. 
              }
         \label{Fig.6}
   \end{figure}


   \begin{figure}
   \centering
   \includegraphics[width=0.47\textwidth]{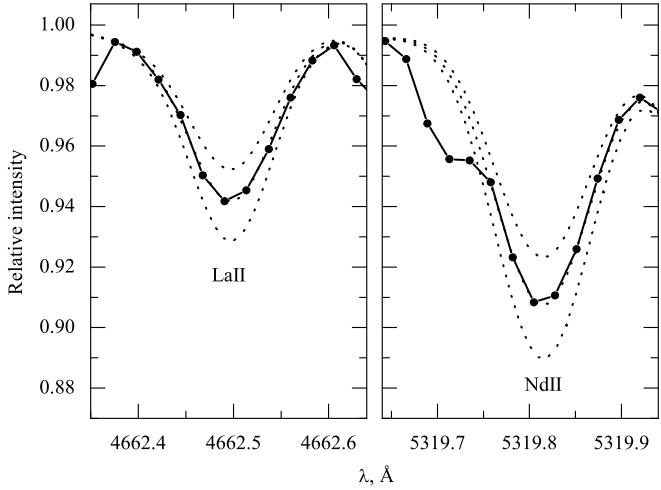}
      \caption{Synthetic spectrum fit to the lanthanum line at 4662~{\AA} in the observed spectrum of HD~143291 (left panel) and to the neodymium line 
      at 5319~{\AA} in the observed spectrum of HD~24156 . The observed spectra are shown by solid lines with dots. The dotted lines are synthetic 
      spectra with ${\rm [La/Fe]} = 0.05 \pm 0.1$ and ${\rm [Nd/Fe]} = 0.12 \pm 0.1$, respectively. 
              }
         \label{Fig.7}
   \end{figure}


   \begin{figure}
   \centering
   \includegraphics[width=0.47\textwidth]{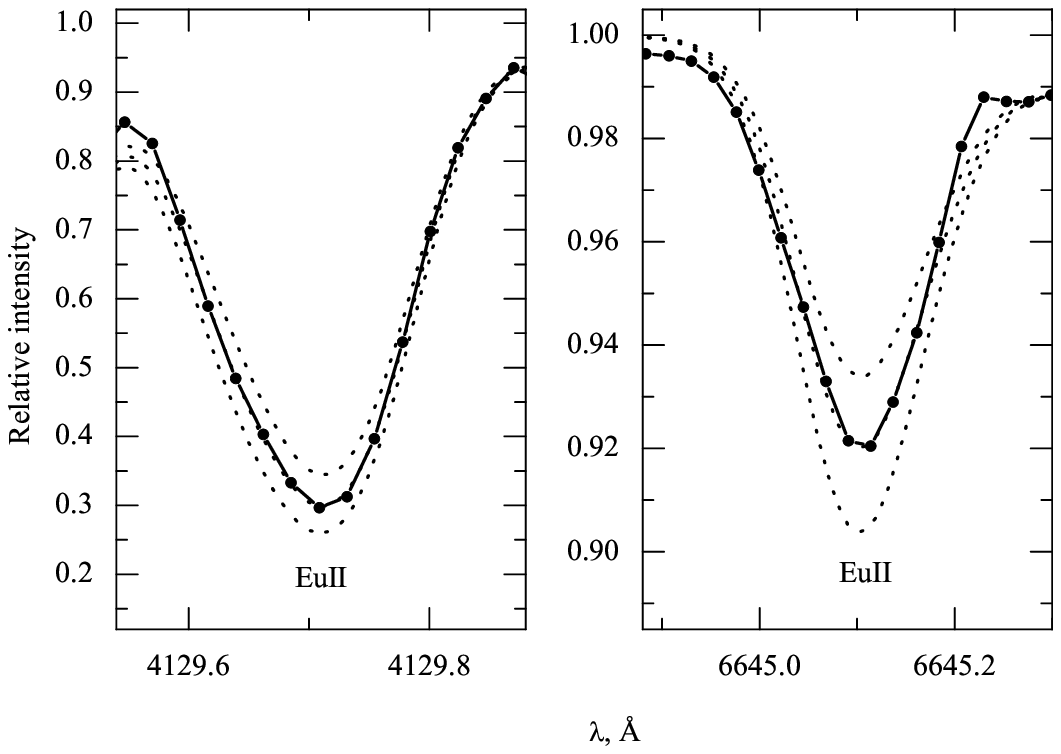}
      \caption{Synthetic spectrum fit to the europium lines at 4129~{\AA} and 6645~{\AA}. The observed spectrum for the programme star 
      HD~12782 is shown as a solid line with dots. The dotted lines are synthetic spectra with ${\rm [Eu/Fe]} = 0.40 \pm 0.1$ and 
      ${\rm [Eu/Fe]} = 0.45 \pm 0.1$,  respectively for these two lines. 
            }
         \label{Fig.8}
   \end{figure}

The Vienna Atomic Line Data Base (VALD, \citealt{piskunov95}) was  
used in preparing input data for the calculations. Atomic oscillator 
strengths for the main spectral lines analysed in this study were taken from  
an inverse solar spectrum analysis performed in Kiev \citep{gurtovenko89}.

Initial values of the effective temperatures for the programme stars were taken from \citet{holmberg09} 
and were then carefully checked and corrected for, if needed, by 
forcing the Fe~{\sc i} lines to yield no dependency of the iron abundance on the excitation 
potential by changing the model effective temperature. 
We used the ionisation equilibrium method to find surface gravities of the programme stars 
by forcing neutral and ionised iron lines to yield the same iron abundances.
Microturbulence velocity values corresponding to the lowest line-to-line  
Fe~{\sc i} abundance scattering were chosen as correct values. 
We determined the abundance of oxygen, yttrium, zirconium, barium, lanthanum, cerium, 
praseodymium, neodymium, samarium and europium with the spectral synthesis method. 
Several fits of the synthetic line profiles to the observed spectra are shown in Figs.~\ref{Fig.5}--\ref{Fig.8}. 
Atomic parameters of lines in the intervals of spectral syntheses were compiled from the VALD database.
All log~\textit{gf} values were calibrated to fit to the solar spectrum by \cite{kurucz05} with solar abundances from \cite{grevesse00}.
Hyperfine structures and isotope shifts were taken into account as appropriate.  
Abundances of other chemical elements were determined using equivalent widths of their lines. 
Abundances of Na and Mg were determined with non-local thermodynamical equilibrium (NLTE) taken into 
account. The equivalent widths of the lines were measured by fitting a Gaussian profile using the {\sc 4A} software 
package \citep{ilyin00}.
  
The uncertainties in abundances are due to several sources: uncertainties caused by 
analysis of individual lines, including random errors of atomic data and continuum 
placement and  uncertainties in the stellar parameters.
The sensitivity of the abundance 
estimates to changes in the atmospheric parameters by the assumed errors 
$\Delta$[El/H]\footnote{We use the customary spectroscopic notation
[X/Y]$\equiv \log_{10}(N_{\rm X}/N_{\rm Y})_{\rm star} -\log_{10}(N_{\rm X}/N_{\rm Y})_\odot$.} 
are illustrated  for the star HD\,10519  (Table~\ref{table:2}). Clearly, possible 
parameter errors do not affect the abundances seriously; the element-to-iron 
ratios, which we use in our discussion, are even less sensitive. 

The scatter of the deduced abundances from different spectral lines $\sigma$
gives an estimate of the uncertainty due to the random errors. The mean value 
of  $\sigma$ is 0.06~dex, thus the uncertainties in the derived abundances that 
are the result of random errors amount to approximately this value.

   \begin{table}
   \centering
   \begin{minipage}{80mm}
      \caption{Effects on derived abundances resulting from model changes for the star HD~10519.} 
        \label{table:2}
      \[
         \begin{tabular}{lrrrr}
            \hline
	    \hline
            \noalign{\smallskip}
Ion & ${ \Delta T_{\rm eff} }\atop{ +100 {\rm~K} }$ & 
            ${ \Delta \log g }\atop{ +0.3 }$ & 
            ${ \Delta v_{\rm t} }\atop{ +0.3~{\rm km~s}^{-1}}$ &
            ${\rm Total} $ \\ 
            \noalign{\smallskip}
            \hline
            \noalign{\smallskip}
[O\,{\sc i}]      & $0.03  $ & $0.12  $ & $0.01  $ & $ 0.12$\\
Na\,{\sc i}       & $0.05  $ & $-0.01 $ & $-0.01 $ & $ 0.05$\\
Mg\,{\sc i}       & $0.05  $ & $-0.02 $ & $-0.03 $ & $ 0.06$\\
Al\,{\sc i}       & $0.03  $ & $-0.02 $ & $-0.02 $ & $ 0.04$\\
Si\,{\sc i}       & $0.03  $ & $0.01  $ & $-0.02 $ & $ 0.04$\\
Ca\,{\sc i}       & $0.07  $ & $-0.01 $ & $-0.03 $ & $ 0.08$\\
Sc\,{\sc ii}      & $0.00  $ & $0.11  $ & $-0.06 $ & $ 0.13$\\
Ti\,{\sc i}       & $0.09  $ & $0.00  $ & $-0.02 $ & $ 0.09$\\
Ti\,{\sc ii}      & $0.01  $ & $0.12  $ & $-0.05 $ & $ 0.13$\\
V\,{\sc i}        & $0.10  $ & $-0.01 $ & $-0.01 $ & $ 0.10$\\
Cr\,{\sc i}       & $0.08  $ & $-0.01 $ & $-0.04 $ & $ 0.09$\\
Fe\,{\sc i}       & $0.07  $ & $-0.01 $ & $-0.06 $ & $ 0.09$\\
Fe\,{\sc ii}      & $-0.02 $ & $0.11  $ & $-0.08 $ & $ 0.14$\\
Co\,{\sc i}       & $0.09  $ & $0.01  $ & $-0.01 $ & $ 0.09$\\
Ni\,{\sc i}       & $0.06  $ & $0.00  $ & $-0.04 $ & $ 0.07$\\
Y\,{\sc ii}       & $0.02  $ & $0.10  $ & $-0.12 $ & $ 0.16$\\
Zr\,{\sc i}       & $0.11  $ & $0.00  $ & $0.01  $ & $ 0.11$\\
Zr\,{\sc ii}      & $0.02  $ & $0.13  $ & $0.01  $ & $ 0.13$\\
Ba\,{\sc ii}      & $0.06  $ & $0.09  $ & $-0.14 $ & $ 0.18$\\
La\,{\sc ii}      & $0.04  $ & $0.12  $ & $0.01  $ & $ 0.13$\\
Ce\,{\sc ii}      & $0.03  $ & $0.12  $ & $0.01  $ & $ 0.12$\\
Pr\,{\sc ii}      & $0.03  $ & $0.12  $ & $0.01  $ & $ 0.12$\\
Nd\,{\sc ii}      & $0.04  $ & $0.12  $ & $-0.01 $ & $ 0.13$\\
Sm\,{\sc ii}	  & $0.04  $ & $0.11  $ & $-0.01 $ & $ 0.12$\\
Eu\,{\sc ii}      & $0.04  $ & $0.12  $ & $0.00  $ & $ 0.13$\\
            \hline
         \end{tabular}
      \]
\end{minipage}
\tablefoot{The table entries show the effects on the logarithmic abundances relative to hydrogen, $\Delta {\rm [El/H]}$. }
   \end{table}

Effective temperatures for all  stars investigated here are also available in
\citet{holmberg09} and \citet{casagrande11}.
\citet{casagrande11} provided astrophysical parameters for the Geneva-Copenhagen survey by applying the infrared flux method to determine the 
effective temperature.
In comparison to \citet{holmberg09}, stars in the catalogue of \citet{casagrande11} are 
on average 100~K hotter. For the stars investigated here, our spectroscopic temperatures are on average only $10\pm 60$~K 
hotter than in \citet{holmberg09} and $60\pm 80$~K cooler than in \citet{casagrande11}.  
[Fe/H] values for all investigated stars are available in \citet{holmberg09} as well as in \citet{casagrande11}. 
A comparison between \citet{holmberg09} and \citet{casagrande11} shows that the latter gives [Fe/H] values that are more metal-rich on average by 
0.1~dex.
For our programme stars we obtain a difference of $0.1\pm 0.1$~dex in comparison with \citet{holmberg09} and no systematic difference, but 
a scatter of 0.1~dex in comparison with \citet{casagrande11}.  The same result was found from comparing the atmospheric parameters determined 
for Group~3 stars in our \citetalias{stonkute12}. 

Some stars from our sample were previously investigated by other authors. In Table~\ref{table:3} we present a comparison with results by
\citet{reddy06}, \citet{mashonkina07}, and \citet{ramirez07}, who investigated several stars in common with our work.
Five thin-disc stars that we investigated in our work for a comparison have been analysed previously by
\citet{edvardsson93}, and two stars have been studied by \citet{bensby05}. 
 Slight differences in the log\,$g$ values lie within the errors of uncertainties and are caused mainly by 
differences in the applied determination methods. In our work we see that titanium and zirconium abundances determined using both neutral and ionised 
lines agree well and confirm the log\,$g$ values determined using iron lines. Overall, our [El/Fe] for the stars in common agree very well with 
those in other studies.

\begin{table}
\centering
\begin{minipage}{80mm}
\caption{Comparison of Group~2 with previous studies.}
\label{table:3} 
\begin{tabular}{lrrrrrrrr}
\hline\hline   
         & \multicolumn{2}{c}{Ours--Reddy} & \multicolumn{2}{c}{Ours--Mashonkina} & \multicolumn{2}{c}{Ours--Ram\'{i}rez} \\
Quantity &   Diff.    & $\sigma$ &  Diff.    & $\sigma$ & Diff.    & $\sigma$\\
\hline
$T_{\rm eff} $  & 118    & 44   & --56    & 29   & 72      & 59 \\
log $g$         & --0.10 & 0.13 & --0.20 & 0.06 & --0.15  & 0.13\\
${\rm [Fe/H]}$  & 0.12 	 & 0.13 & 0.04 	 & 0.04 & 0.08    & 0.05 \\
${\rm[Na/Fe]}$  & --0.07 & 0.03 & ...    & ... &  ...    & ...\\
${\rm[Mg/Fe]}$  & 0.01	 & 0.06 &  ...    & ...  & ...    & ...\\
${\rm[Al/Fe]}$  & --0.02 & 0.08 &  ...    & ...  & ...    & ...\\
${\rm[Si/Fe]}$  & 0.01 	 & 0.06 &  ...    & ...  & ...    & ...\\
${\rm[Ca/Fe]}$  & 0.05	 & 0.04 &  ...    & ...  & ...    & ...\\
${\rm[Sc/Fe]}$  & 0.07   & 0.11 &  ...    & ...  & ...    & ...\\
${\rm[Ti/Fe]}$  & 0.06	 & 0.08 &  ...    & ...  & ...    & ...\\
${\rm[V/Fe]}$   & 0.01   & 0.15 &  ...    & ...  & ...    & ...\\
${\rm[Cr/Fe]}$  & 0.02	 & 0.04 &  ...    & ...  & ...    & ...\\
${\rm[Co/Fe]}$  & 0.00	 & 0.04 &  ...    & ...  & ...    & ...\\
${\rm[Ni/Fe]}$  & 0.00 	 & 0.05 &  ...    & ...  & ...    & ...\\
${\rm[Y/Fe]}$	& --0.06 & 0.08 & --0.10 & 0.08 & ...     & ... \\
${\rm[Zr/Fe]}$	& ...    & ...  & --0.07 & 0.08 & ...    & ...  \\
${\rm[Ba/Fe]}$	& 0.08   & 0.08 & 0.01   & 0.06 & ...     & ... \\
${\rm[Ce/Fe]}$	& --0.04 & 0.16 & --0.11 & 0.11 & ...     & ... \\
${\rm[Nd/Fe]}$	& --0.12 & 0.23 & ...    & ...  & ...     & ... \\
${\rm[Eu/Fe]}$	& 0.03   & 0.09 & ...    & ...  & ...     & ... \\
\hline
\end{tabular}
\end{minipage}
\tablefoot{Mean differences and standard deviations of the main parameters and abundance ratios [El/Fe] for
 7 stars in common with \citet{reddy06}, 5 stars in common with \citet{mashonkina07}, and 10 stars in common with \citet{ramirez07}.}
\end{table}

\section{Results and discussion}

The atmospheric parameters $T_{\rm eff}$, log\,$g$, $v_{t}$, [Fe/H] and abundances of 21 chemical elements relative 
to iron [El/Fe] of the programme and comparison stars are presented in Tables~\ref{table:4} and \ref{table:5}. The number of lines and 
the line-to-line scatter ($\sigma$) are presented as well. 

The metallicities of Group~2 stars lie in quite a narrow interval, with a mean of ${\rm [Fe/H]}= -0.42\pm 0.10$. Abundances of other chemical 
elements are quite homogeneous and show similar overabundances of $\alpha$-elements and r-process-dominated chemical elements with 
respect to thin-disc stars, as was also found for the stars of CGS Group~3.  However, two stars (HD~200580 and HD~224817) have abundances 
of these elements similar to the thin-disc stars of the same metallicity. According to their element-to-iron ratios, these two stars might not belong to  Group~2. However, their ages ($\sim 8$ and 12~Gyr, respectively) as presented in Subsect.~3.1 seem pretty
old and similar to those of other member stars of Group 2. Their kinematic parameters are also similar to those of other stars in Group~2, therefore 
we have only one argument for detracting a Group~2 membership from these two stars.    

  \begin{figure*}
   \centering
   \includegraphics[width=0.85\textwidth]{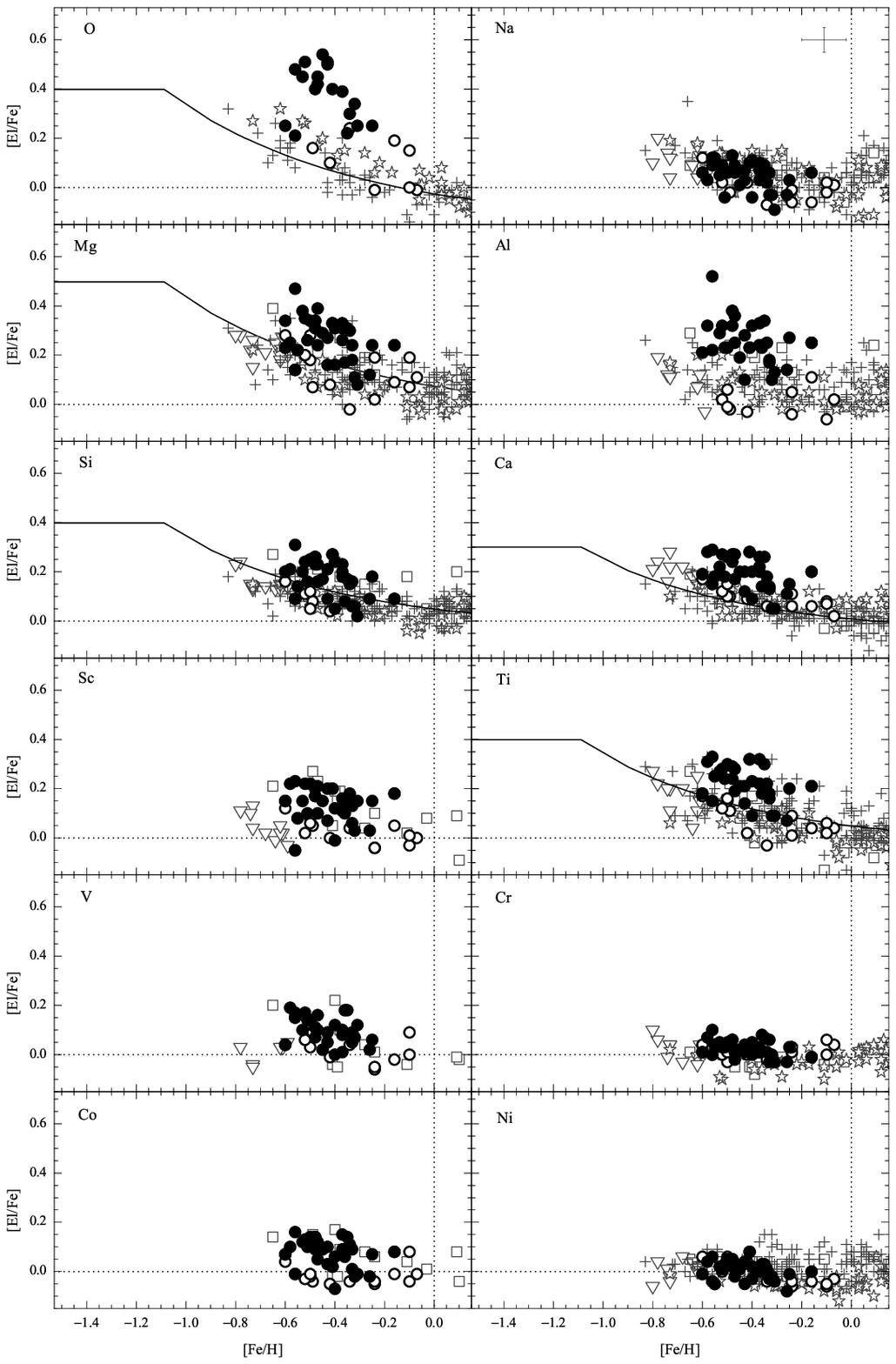}
     \caption{[El/Fe] ratio as a function of [Fe/H] in the investigated stars of Group 2 (filled circles) and comparison stars analysed in this work and 
     \citetalias{stonkute12} (open circles). The data for the Milky Way thin-disc dwarfs were taken from \citeauthor{edvardsson93} (1993, plus signs), 
     \citeauthor{bensby05} (2005, stars), \citeauthor{reddy06} (2006, squares), and \citeauthor{zhang06} (2006, triangles). Solid lines are Galactic thin-disc 
     chemical evolution models presented by \citeauthor{pagel95} (1995). Average uncertainties are shown in the box for Na.}
       \label{Fig.9}
   \end{figure*}
   
  \begin{figure*}
   \centering
   \includegraphics[width=0.85\textwidth]{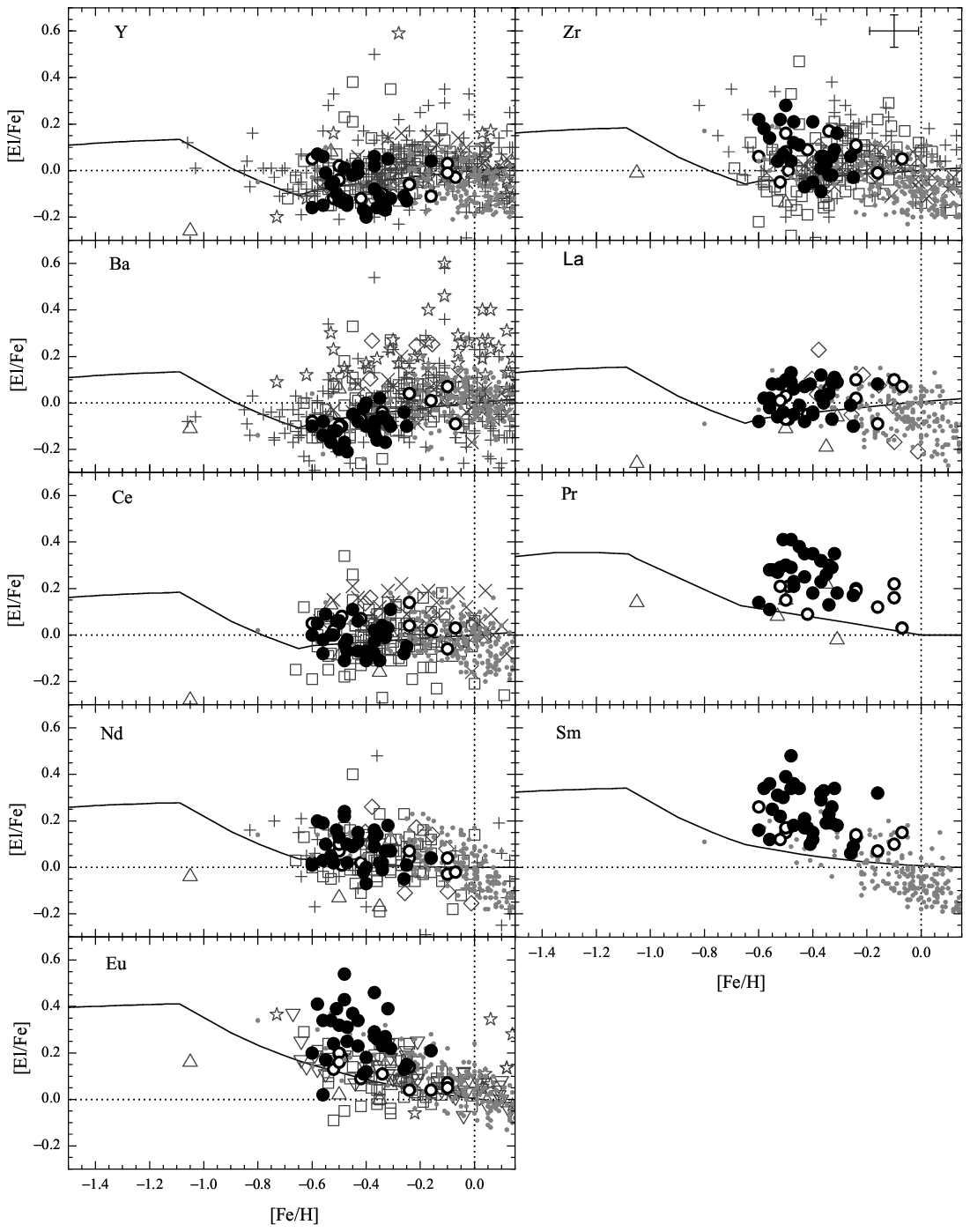}
     \caption{[El/Fe] ratio as a function of [Fe/H] in the investigated stars of Group 2 (filled circles) and comparison stars analysed in this work and 
     \citetalias{stonkute13} (open circles). The data for the Milky Way thin-disc dwarfs were taken from \citeauthor{edvardsson93} (1993, plus signs), 
     \citeauthor{gratton94} (1994, triangles), \citeauthor{koch02} (2002, upside down triangles), \citeauthor{bensby05} (2005, stars), 
     \citeauthor{reddy06} (2006, 2003, squares), \citeauthor{brewer06} (2006, diamonds), \citeauthor{mashonkina07} (2007, crosses), and
     \citeauthor{mishenina13} (2013, dots). The Galactic thin-disc chemical evolution model is shown as a solid line (\citeauthor{pagel97}, 1997). 
     Average uncertainties are shown in the box for Zr.}
       \label{Fig.10}
   \end{figure*}

  \begin{figure*}
   \centering
   \includegraphics[width=0.85\textwidth]{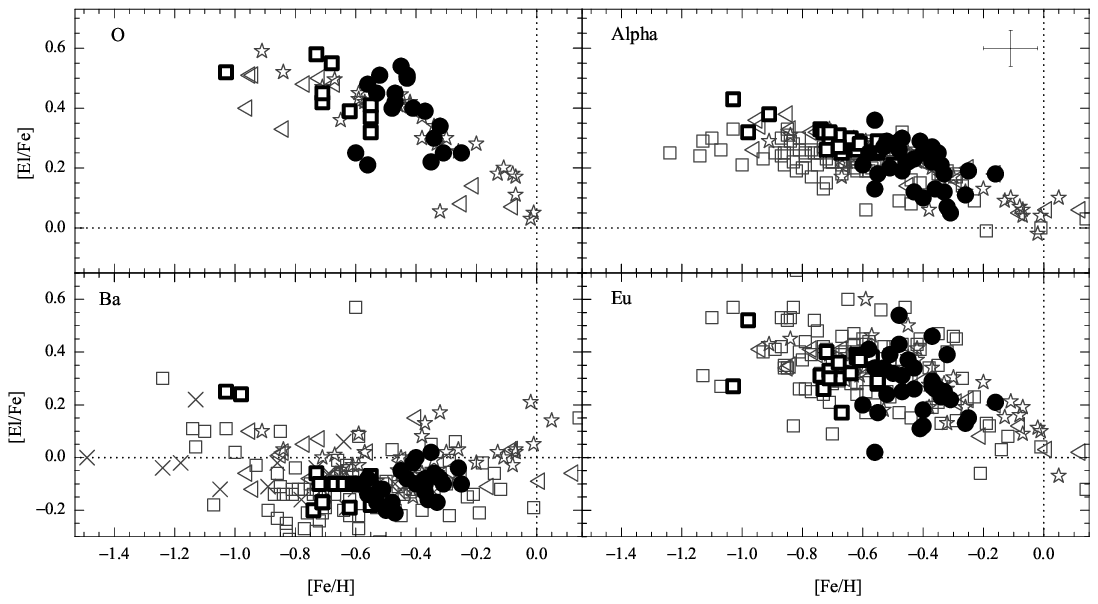}
     \caption{[El/Fe] ratio as a function of [Fe/H] in the investigated stars of Group 2 (filled circles) and Group~3 (thick squares, Papers~I and II). 
     The data for the Milky Way thick-disc stars were taken from \citeauthor{mishenina13} (2013, triangles), \citeauthor{mashonkina07} (2007, crosses), 
     \citeauthor{reddy06} (2006, squares), and \citeauthor{bensby05} (2005, stars).  The averaged values for alpha elements consist of Mg, Si, and Ca 
     abundances.}
       \label{Fig.11}
   \end{figure*}

The results are graphically displayed in Figs.~\ref{Fig.9} and \ref{Fig.10}.  We display elemental abundance ratios of Group~2 stars together with 
data of thin-disc stars investigated in this work and in Papers~I and II, as well as with results taken from thin-disc studies (\citealt{edvardsson93, 
gratton94, koch02, bensby05, reddy06, zhang06, brewer06, mashonkina07}, and \citealt{mishenina13}). The chemical evolution models of the 
thin-disc were taken from \citet{pagel95, pagel97}. 
The thin-disc stars from \citet{edvardsson93} and \citet{zhang06} were selected by using the membership probability evaluation method 
described by \citet{trevisan11},
since their lists contained stars of other Galactic components as well. The same kinematical approach in assigning thin-disc membership was used in  
\citet{bensby05} and \citet{reddy06}, which means that the thin-disc stars used for the comparison are uniform in that respect. 
   
Abundances of $\alpha$-elements and r-process-dominated elements (particularly of Eu, Sm, and Pr) in the investigated Group~2 stars are 
higher than those of the Galactic thin-disc dwarfs investigated in this work and other studies. We found similar overabundances at the 
corresponding metallicity in stars of Group~3 as well. 

This pattern of elemental ratios is exhibited by the thick-disc stars. In Table~\ref{table:6} we present a comparison of mean [El/Fe] ratios 
calculated for stars of Group 2 and thick-disc stars at the same metallicity interval $-0.6 < {\rm [Fe/H]} < -0.2$. Twenty-one stars in this 
metallicity interval were investigated by  \citet{bensby05}, 38 stars by \citet{reddy06}, 11 stars by \citet{mashonkina07}, 
and 7 stars by \citet{mishenina13}. When comparing oxygen abundances, we did not use the results reported by \citet{reddy06} because they 
investigated the O\,{\sc i}  line, while we studied [O\,{\sc i}] and neither did we use the results reported by \citet{mishenina13} because only three thick-disc stars fall in the corresponding metallicity 
interval. The studies by \citet{mashonkina07} and \citet{mishenina13} were included in the comparison to enlarge the information 
on neutron capture elements.  The average values of $\alpha$-element abundances included Mg, Si, and Ca. Titanium was excluded because this element was not determined in one of the studies (\citealt{mishenina13}). The comparison shows that the 
deviations do not exceed the uncertainties. 
 
 \begin{table}
 \setcounter{table}{5}
   \centering
   \begin{minipage}{80mm}
      \caption{Comparison with thick-disc studies.} 
        \label{table:6}
      \[
         \begin{tabular}{lrrrr}
            \hline
	    \hline
            \noalign{\smallskip}
[El/Fe] & ${\rm Ours-}\atop {\rm Bensby}$ & 
            ${\rm Ours- }\atop{\rm Reddy }$ & 
            ${\rm Ours- }\atop{\rm Mashonkina}$ &
            ${\rm Ours- }\atop{\rm Mishenina} $ \\ 
            \noalign{\smallskip}
            \hline
            \noalign{\smallskip}
${\rm[O/Fe]}$        & $0.04   $ & $...   $ & $...   $ & $ ...$\\
${\rm[Na/Fe]}$		& $-0.03   $ & $-0.03   $ & $...   $ & $ ...$\\
${\rm[Mg/Fe]}$       & $-0.01  $ & $0.00  $ & $...   $ & $ 0.02$\\
${\rm[Al/Fe]}$		& $-0.01   $ & $0.01   $ & $...   $ & $ ...$\\
${\rm[Si/Fe]}$       & $0.00   $ & $-0.02 $ & $...   $ & $ 0.00$\\
${\rm[Ca/Fe]}$       & $0.03   $ & $0.05  $ & $...   $ & $ 0.02$\\
${\rm[Sc/Fe]}$		& $...   $ & $-0.01   $ & $...   $ & $ ...$\\
${\rm[Ti/Fe]}$       & $0.01   $ & $0.04  $ & $...   $ & $ ...$\\
${\rm[V/Fe]}$		& $...   $ & $-0.02   $ & $...   $ & $ ...$\\
${\rm[Cr/Fe]}$		& $0.02   $ & $0.04   $ & $...   $ & $ ...$\\
${\rm[Co/Fe]}$		& $...   $ & $-0.02   $ & $...   $ & $ ...$\\
${\rm[Ni/Fe]}$		& $-0.01   $ & $-0.02   $ & $...   $ & $ -0.02$\\
${\rm[Y/Fe]}$        & $-0.05  $ & $-0.09 $ & $-0.11 $ & $ -0.07$\\
${\rm[Zr/Fe]}$       & $...    $ & $...   $ & $-0.03 $ & $ 0.01$\\
${\rm[Ba/Fe]}$       & $-0.13  $ & $0.00  $ & $0.01  $ & $ -0.04$\\
${\rm[La/Fe]}$       & $...    $ & $...   $ & $...   $ & $ 0.07$\\
${\rm[Ce/Fe]}$       & $...    $ & $-0.09 $ & $-0.06 $ & $ 0.02$\\
${\rm[Nd/Fe]}$       & $...    $ & $-0.10 $ & $...   $ & $ -0.02$\\
${\rm[Sm/Fe]}$       & $...    $ & $...   $ & $...   $ & $ 0.10$\\
${\rm[Eu/Fe]}$       & $-0.05  $ & $-0.05 $ & $...   $ & $ 0.03$\\
            \hline
         \end{tabular}
      \]
\end{minipage}
\tablefoot{Differences of mean [El/Fe] values for stars of Group 2 and thick-disc stars at the same metallicity interval $-0.6 <$ [Fe/H] $< -0.2$. 
21 stars from \citet{bensby05}, 38 stars from \citet{reddy06}, 11 stars from \citet{mashonkina07}, and 7 stars from 
\citet{mishenina13}. }
   \end{table}
   
Fig.~\ref{Fig.11} displays the comparison of [El/Fe] ratios for some chemical elements between individual stars in Groups~2 and 3 and the 
thick-disc stars of the above-mentioned studies. For the comparison we selected oxygen, the averaged values for the alpha elements Mg, Si, and 
Ca, and the s- and r-process-dominated elements barium and europium, respectively. Stars of the kinematic groups and of the thick-disc have a 
very similar chemical composition. To reveal some possible tiny differences, both types of stars should be investigated using an identical method. 
However, as we outlined in \citetalias{stonkute12} for Group~3, the similar chemical composition of stars in Group~2 and the thick-disc stars 
might also suggest that their formation histories are linked.

\subsection{Age}

Group~2 is characterised by an interesting feature related to the age distribution of its stars.   According to \citet{helmi06}, the stars fall into three 
populations: 15\% of the stars are 8~Gyr old, 36\% are 12~Gyr old,  and 49\% are 16~Gyr old. The ages were later redetermined by \citet{holmberg09} and  
\citet{casagrande11}. Because we redetermined the effective temperatures and metallicities using high-resolution spectra, we revisited the age determinations as well.  

For the age determination we used the method by \citet{jorgensen05} which was used previously in CGS studies (\citealt{helmi06, holmberg09}).  This method and 
other similar Bayesian methods are currently the most common way to determine ages for larger stellar samples. This method with Bayesian 
probability functions gives a better understanding of how accurate the ages are. For the age determination of Group~2 stars we took into account 
the $\alpha$/Fe overabundance of 0.2~dex. The new age evaluations together with lower and upper age limits are presented in Table~\ref{table:7}. 
Previously determined ages from \citet{holmberg09} and \citet{casagrande11} are presented as well. 

\begin{table}
	\centering
\begin{minipage}{80mm}
\caption{Ages determined in this work and other studies. }
\label{table:7}
      \[
\begin{tabular}{lrrrrr}
\hline  
\hline
         Star &  H09  &	 C11  & This work&  $-1\sigma$ &  $+1\sigma$  \\
\hline
\noalign{\smallskip}
{BD +68 813}  & 13.3  & 9.2   & 9.8	&	7.5     & 11.8 \\        
{BD +31 3330} & ...   & 7.1   &	...	&	...     & ... \\      
{HD 10519}    & 12.4  & 11.8  & 12.0	&   10.9    & 13.1 \\        
{HD 12782}    &	...& ...   &    ...   &	...     & ...  \\      
{HD 16397}    & 10.6  & ...   & 10.4	&	 7.9    & 12.4	\\    
{HD 18757}    & 16.8  & ...   & 11.0  &	 8.9    & 16.2	\\    
{HD 21543}    & 11.6  & 8.3   & 13.5  &	 9.3    & 17.4  \\     
{HD 24156}    & 11    & 10.2  & 10.4  &	 8.9    & 11.8 	\\    
{HD 29587}    & 11.7  & 8.5   & 8.3   &	 2.1    & 13.3 	\\    
{HD 30649}    & 10.1  & 7.8   & 9.8   &	 8.1    & 12.5 	\\    
{HD 37739}    & 3.4   & 3.4   & 3.3   &	 3.0    &  3.7 	\\    
{HD 38767}    & 4.9   & ...   & 5.0   &	 3.9    &  5.8 \\     
{HD 96094}    & 8.6   & 8.7   & 6.3   &	 4.9    &  7.9 	\\    
{HD 114606}   & 16.6  & 7.6   & 15.1  & 10.8	 & 17.7 	\\    
{HD 121533}   & 12.6  & ...   & 14   &	 10.3   & 17.7  	\\    
{HD 131582}   & ...   & ...   &	...     &	 ...    & ...  \\      
{HD 132142}   & ...   & 6.9   &	3.5	&	 0.1   & 17.7  \\      
{HD 133621}   & 11.8  & 10.5  & 10.9  &	10.0    & 11.7 	\\    
{HD 137687}   & ...   & 7.1   & 8.7   &	 7.9    &  9.9 	 \\    
{HD 139457}   & 7.7   & 7.1   & 7.3   &	 5.8    &  8.0  	\\    
{HD 143291}   & ...   & 6.1   &	...	&	 ...    &  ...   \\     
{HD 152123}   & 3.6   & 3.4   & 2.5   &	 2.3    &  2.8 	\\    
{HD 156802}   & 9.7   & 7.9   & 9.1   &	 7.3    & 11.3       \\
{HD 158226}   & 12.6  & 8.1   & 11.8  &	10.4    & 13.5  \\     
{HD 165401}   & 8     & 5.1   & 2.5   &	 0.1    &  7.5   \\     
{HD 170357}   & 12.7  & 8.5   & 10.0  &	 8.9    & 11.0 	\\    
{HD 190404}   & 0.2   & ...   &	5	&	 0.1   & 17.7  \\      
{HD 200580}   & 8     & 8.4   & 8.4   &	 7.6    &  9.4 	\\    
{HD 201099}   & 7.5   & 7.1   & 7.2   &	 6.2    &  8.5   	\\    
{HD 215594}   & 10.2  & ...   & 7.8   &	 6.4    &  9.5  	\\    
{HD 221830}   & 12.4  & 9.3   & 11.5  &	10.2    & 12.9 	\\    
{HD 224817}   & 10.3  & 9.5   & 12.0  &	10.5    & 13.9 	 \\    
  \hline
\end{tabular}
      \]
\end{minipage}
\tablefoot{H09 -- ages taken from \citet{holmberg09}, C11 -- from \citet{casagrande11}, ages determined in this work together with lower and upper 
age limits are presented in the last three columns. All ages are in Gyr.}
\end{table}

\begin{figure*}
   \centering
   \includegraphics[width=0.80\textwidth]{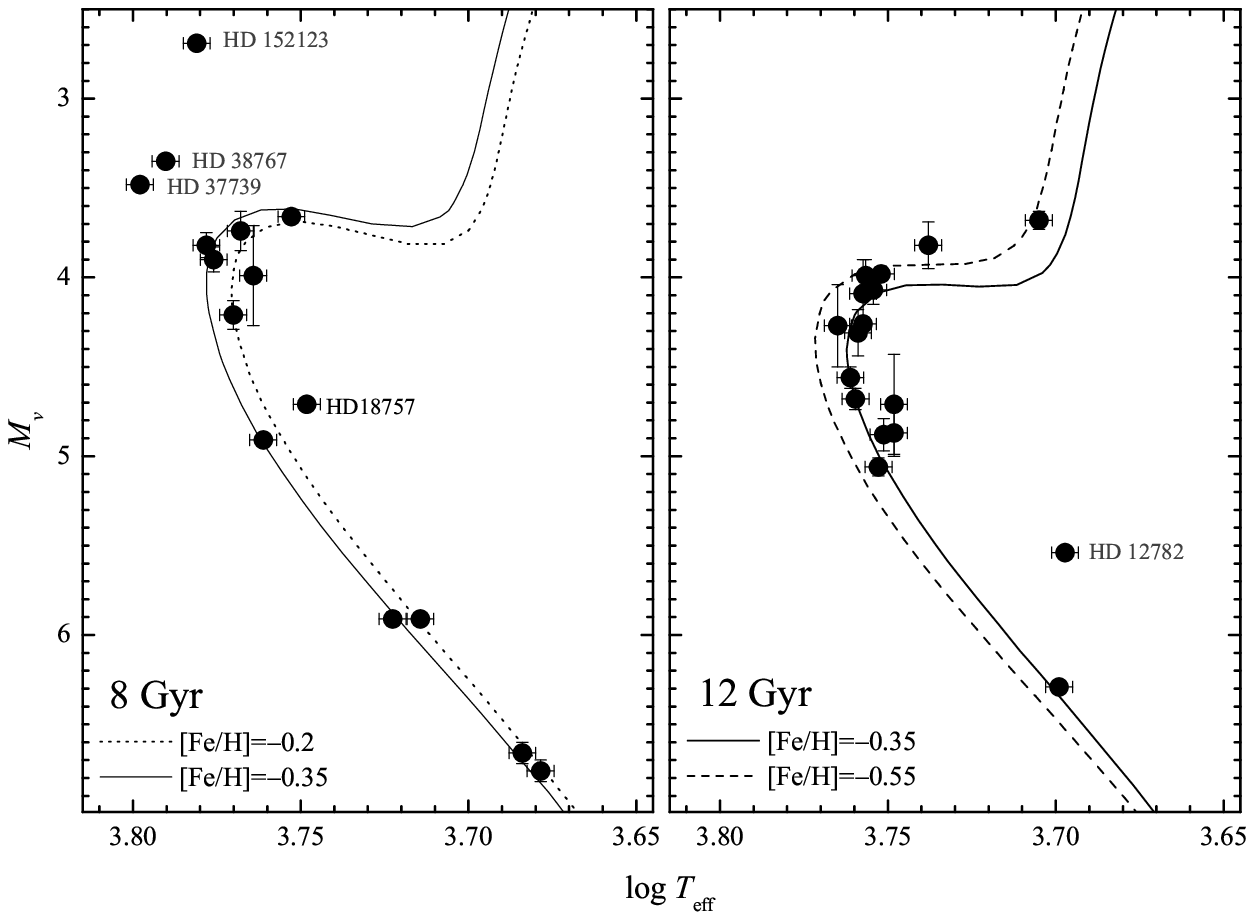}
     \caption{HR diagrams of the Group 2 stars. Isochrones are taken from \citet{bressan12}. The filled circles 
     correspond to the investigated stars with the spectroscopic effective temperatures. Isochrones are with [$\alpha$/Fe ]= 0.2.}
       \label{Fig.12}
   \end{figure*}

Fig.~\ref{Fig.12} shows the investigated stars 
of Group 2 with our spectroscopic effective temperatures and absolute magnitudes $M_{v}$, taken from \citet{holmberg09}, in a Hertzsprung-Russell 
(HR) diagram. The isochrones enhanced by $\alpha/{\rm Fe}\sim 0.2$ were taken from  \citet{bressan12}. 
The overall features of stars in the diagram are well reproduced by isochrones of two ages. The more metal-abundant 
stars fit the 8~Gyr isochrone quite well, while more metal-deficient stars fit the 12~Gyr isochrone.  
Most of the stars for which older or younger ages 
were determined also fit these two age populations because these stars belong to the main sequence and determining their age accurately is problematic.  As we can see from Fig.~12, the stars HD~37739, HD~38767, and HD~152123 are certainly 
younger ($2.5~{\rm Gyr} \leq {\rm age} \leq 5~{\rm Gyr}$). In the HR diagram, they lie higher than the turn-off luminosity of the 8~Gyr isochrone.  
It seems that a subgroup of about 15 such young main sequence stars can be separated from 86 Group~2 stars. A detailed chemical composition 
study of all these stars might be useful. A chemical composition pattern of these young stars investigated in our work is similar to the rest of the
Group~2 stars.             

  \begin{figure*}
   \centering
   \includegraphics[width=0.90\textwidth]{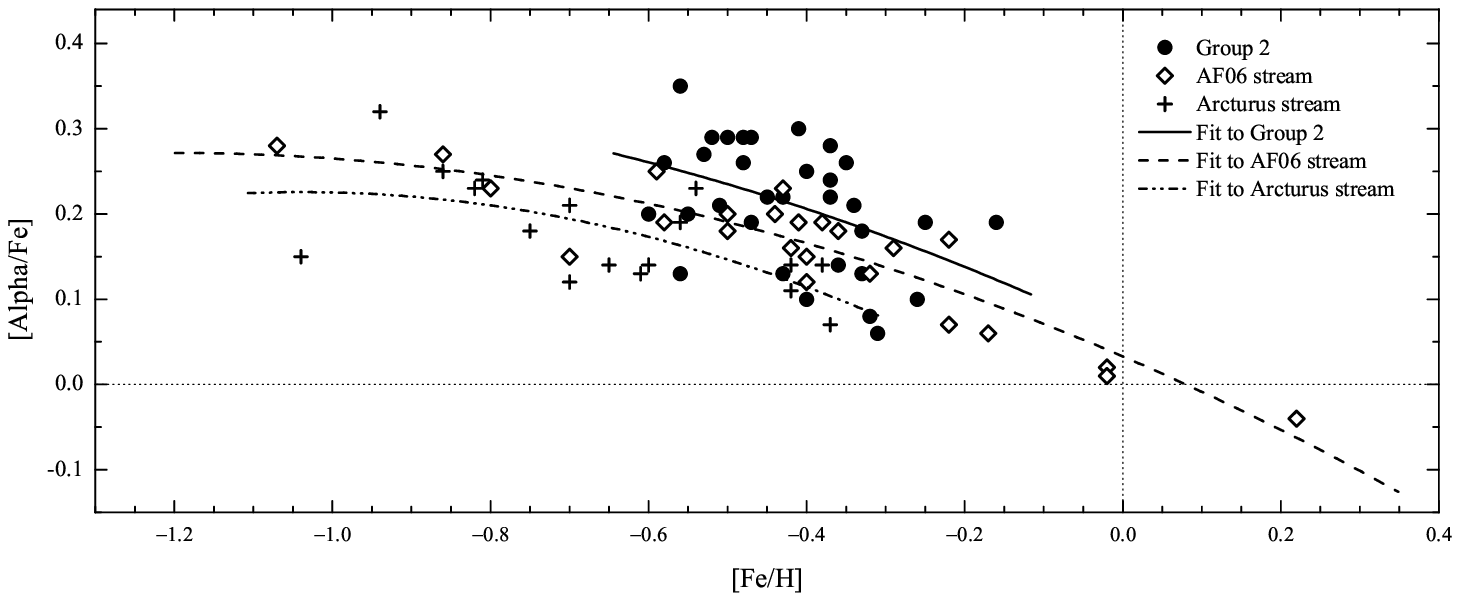}
     \caption{[Alpha/Fe] ratio as a function of [Fe/H] in the investigated stars of Group 2 (filled circles), in the Arcturus stream (plus signs), and in the 
     AF06 stream (diamonds). The data for Arcturus and AF06 streams were taken from \citet{ramya12}.   The averaged values for alpha elements 
     consist of Mg, Si, Ca, and Ti abundances.}
       \label{Fig.13}
   \end{figure*}

\subsection{Origin}
 
The similar chemical composition pattern of stars in GCS Groups~2 and 3 (Papers~I and II) indicates that these kinematic groups share 
a similar origin.   The similarity in chemical composition of stars in these two kinematic groups and in the thick-disc of the Galaxy suggests 
that the formation histories of these groups and the thick-disc might be linked as well.  Thus it is natural to determine which of the currently available thick-disc  
formation scenarios is most suitable. 
The kinematic properties of Groups~2 and 3 fit a gas-rich satellite merger scenario well (\citealt{brook04}, 2005; \citealt{dierickx10,
wilson11, dimatteo11}, and references therein).  Within this specific scenario, the eccentricities of accreted stars peak at about 
$ 0.3 < \epsilon < 0.5$ (\citealt{sales09}), which is exactly the characteristics of the investigated CGS groups. This scenario fits the 
thick-disc star eccentricity distribution better than the accretion, heating, or migration scenarios (\citealt{dierickx10}).    

\citet{dierickx10} analysed the eccentricity distribution of thick-disc stars that has recently been proposed as a diagnostic to 
differentiate between these mechanisms \citep{sales09}. Using SDSS data release 7, they have assembled a sample 
of 31.535 G-dwarfs with six-dimensional phase-space information and metallicities and have derived their orbital eccentricities. 
They found that the observed eccentricity distribution is inconsistent with that predicted by orbital migration alone. 
Moreover, the thick-disc cannot be produced predominantly through heating of a pre-existing thin-disc, since this model predicts more high-eccentricity 
stars than observed. According to \citet{dierickx10}, the observed eccentricity distribution  fits a gas-rich merger scenario well, where 
most thick-disc stars were born \textit{in situ}. 

In the gas-rich satellite merger scenario, a distribution of stellar eccentricities peak around $\epsilon=0.25$, with a tail towards higher 
values belonging mostly to stars originally formed in satellite galaxies. The group of stars investigated in our work fits this model 
with a mean eccentricity value of 0.4. This scenario is also supported by the RAVE survey data analysis made by 
\citet{wilson11} and the numerical simulations by \citet{dimatteo11}. In this scenario, Groups~2 and 3 can be explained as 
remnants of stars originally formed in a merging satellite. These kinematic star groups might belong to the same satellite galaxy. 
\citet{helmi06} showed that in numerical simulations of the disruption of a satellite galaxy that falls into its parent galaxy, the satellite debris 
can end up in several cold star streams with roughly the same characteristic eccentricities of their orbits. To conclude about the origin of the kinematic CGS groups, we need more observational and theoretical investigations of stellar kinematic groups and the thick-disc. In the 
following subsection we briefly overview the available high-resolution spectral analyses of two other kinematic groups of the Galaxy.     

\subsection{Comparison with the Arcturus and AF06 streams}

Finally, we discuss Group~2 in the context of several other Galactic substructures.  
\citet{helmi06}  briefly mentioned a seeming  
overlap between Group 2 and the Arcturus stream (\citealt{navarro04}) in their kinematics and metallicities. Our study of Group~2 and the high-resolution spectroscopic analysis available for the Arcturus stream (\citealt{ramya12}) allow us to compare the detailed chemical composition 
in these two substructures. To this comparison, we added the so-called stellar stream AF06, which was discovered 
by \citet{arifyanto06}. According to the discoverers,  AF06 possibly resembles the Arcturus stream and probably has an extragalactic origin.
A chemical composition of AF06 stars was also investigated by \citet{ramya12}. 

Fig.~\ref{Fig.13} displays the comparison of [El/Fe] ratios for alpha-elements between individual stars in Group~2 and the Arcturus and AF06 streams. 
The averaged values for alpha-elements consist of Mg, Si, Ca, and Ti abundances. 
Simple second-order polynomial fits were applied to the available 18 stars of the Artcturus stream, to 26 stars of the AF06 stream, and to 32 stars of 
Group~2, which revealed slight systematic differences between the element-to-iron ratios in these stellar groups.  While Group~2 and AF06 stars differ 
only by about 0.05~dex, the Arcturus stars lie by about 0.1~dex lower than the Group~2 stars.  Consequently, the detailed 
chemical composition of the Group~2 and Arcturus stream stars is different. We disagree with \citet{ramya12}, who reported that the Arcturus group overlaps  
with the properties of the local field thick-disc stars. In our understanding,  the chemical composition of stars attributed to the Arcturus stream, 
determined in the work by \citet{ramya12}, much more resembles the abundances of the thin-disc than those of the thick-disc stars. 

The origin of the Arcturus stream has been debated for years.  The identification of the Arcturus stream member stars has begun in 1971 and still continues (\citealt{eggen71}, 1996, 
1998; \citealt{arifyanto06, gilmore02, wyse06, bensby13}, and references therein). 
\citet{navarro04} analysed the group of stars associated kinematically with Arcturus and confirmed that they constitute a peculiar grouping 
of metal-poor stars with a similar apocentric radius, a common angular momentum, and distinct metal abundance patterns. These properties are 
consistent with those expected for a group of stars originating from the debris of a disrupted satellite. It was also noticed that its angular momentum 
appears to be too low to arise from dynamical perturbations induced by the Galactic bar. 
More recently, \citet{gardner10} and \citet{monari13} showed that the Galactic long bar may produce a kinematic feature in velocity space 
with the same parameters as occupied by the Arcturus moving group. However, if the Arcturus group indeed has thick-disc kinematics and thin-disc 
abundances, its peculiarity and possible extragalactic origin remain an unsolved question.   

For the AF06 stream, our comparison shows that the chemical composition of this stream is quite similar to that of Group~2.  The AF06 stream was 
identified by \citet{arifyanto06} analysing the fine structure of the phase space distribution function of nearby subdwarfs using data extracted 
from various catalogues. 

It is worth pointing out that currently observed Galactic satellites, which are often referred to as representatives of the building blocks of our Galaxy, 
contain systematically lower $[\alpha/{\rm Fe}]$ ratios at corresponding metallicities than observed in these kinematic 
groups as well as in field stars of the Galaxy, and stars in each dwarf galaxy also show different abundance patterns (see e.g. \citealt{geisler07, tautvaisiene09, tolstoy09}, and references therein). 
Moreover, photometric and spectroscopic studies of dwarf spheroidal galaxies show that they contain little or no gas and no recent star formation 
(e.g. \citealt{smecker94, tolstoy03, venn04}). Various solutions have been proposed to solve these problems  
(\citealt{robertson05, bullock05, font06}a,b; \citealt{kirby08}, 2011; \citealt{frebel10, tafelmeyer10, smith12, belokurov13}), it was even proposed that the present-day Local Group dwarf galaxies may not be the generic galactic building 
blocks (\citealt{unavane96}). This means that there is certainly much more to discover about our Milky Way and studies should be continued. We plan to analyse 
a chemical composition of stars in Group~1 of the Geneva-Copenhagen survey. This group is the most metal-abundant of the new kinematic 
groups identified in CGS.

\section{Conclusions}

We measured abundances of 22 chemical elements from high-resolution spectra 
in 32 stars belonging to Group~2 of the Geneva-Copenhagen survey. This kinematically identified group of stars 
was suggested to be a remnant of a disrupted satellite galaxy.
Our main goal was to investigate the chemical composition of the stars within the group and to compare them with Galactic disc stars. 
         
Our study of 32 stars in Group~2 shows the following: 
  \begin{enumerate}
     \item The metallicities of the investigated stars in Group~2 are in the range $-0.16 \geq {\rm [Fe/H]} \geq -0.60$. The average 
     [Fe/H] value is $-0.42\pm 0.10$~dex.
    \item  All programme stars have higher abundances in oxygen, $\alpha$-elements, and r-process-dominated chemical elements than Galactic 
    thin-disc dwarfs and the Galactic evolution model. This abundance pattern has similar characteristics as the 
    Galactic thick disc and Group~3 of the Geneva-Copenhagen Survey.  

     \item The abundances of iron-group chemical elements and elements produced mainly by the s-process are similar to those in the Galactic 
     thin-disc dwarfs of the same metallicity. 
       
    \item The similarity of the chemical composition in Group~2 and thick-disc stars  might suggest that their formation histories are linked. 
    
    \item Groups~2 and 3 might originate from the same satellite galaxy.   

     \item Investigated Group~2 stars consist mainly of two 8- and 12-Gyr-old populations.  
 
     \item The chemical composition together with the kinematic properties and ages of stars in the 
investigated Group~2 of the Geneva-Copenhagen survey support a gas-rich satellite merger scenario as the most 
probable origin for Group~2. 
 
  \end{enumerate} 

\begin{acknowledgements}
The data are based on observations made with the Nordic Optical Telescope, operated on the island of 
La Palma jointly by Denmark, Finland, Iceland, Norway, and Sweden, 
in the Spanish Observatorio del Roque de los Muchachos of the Instituto de Astrofisica de Canarias.  
The research leading to these results has received funding from the European Community's 
Seventh Framework Programme (FP7/2007-2013) under
grant agreement number RG226604 (OPTICON). BN acknowledges support from the Danish Research council and the Carlsberg Foundation. 
We are grateful to J. Holmberg for help with age calculations. 
This research has made use of Simbad, VALD and NASA ADS databases. 
We thank the anonymous referee for insightful questions and comments.
\end{acknowledgements}

\Online
\appendix

{\scriptsize
\longtab{4}{
\begin{longtable}{lccccccccccccccc}
\caption{Main atmospheric parameters and elemental abundances of the programme and comparison stars.\label{table:4}} \\

\hline\hline   
Star & $T_{\rm eff}$ & log~$g$ & $v_{t}$ & [Fe/H] & $\sigma_{\rm Fe I}$ & ${\rm n}_{\rm Fe I}$ & $\sigma_{\rm Fe II}$ & ${\rm n}_{\rm Fe II}$& [O/Fe]& [Na/Fe]& $\sigma$& n& [Mg/Fe]& $\sigma$& n\\
   & K             &       & km s$^{-1}$   &      &      &    &    &  \\
 \hline
 \endfirsthead
\caption{continued.}\\
\endhead
 BD +68 813	& 5820	& 4.4	& 1.2 	& --0.48 & 0.05	& 29 & 0.05 & 6 & ...  	& 0.07	&	0.03 &	3 & 0.31 & 0.05	& 4 \\
 BD +31 3330	& 4830	& 4.4	& 0.8	& --0.32 & 0.03	& 31 & 0.04 & 5 & 0.34	& --0.03 &	0.07 &	4 & 0.11 & 0.06	& 4 \\
 HD 10519	& 5680	& 3.8	& 0.9	& --0.52 & 0.04	& 33 & 0.05 & 7 & 0.51	& 0.09	&	0.03 &	3 & 0.35 & 0.01	& 4 \\
 HD 12782	& 4980	& 3.4	& 1.1	& --0.48 & 0.04	& 35 & 0.03 & 7 & 0.40	& 0.13	&	0.03 &	3 & 0.34 & 0.04	& 4 \\
 HD 16397	& 5750	& 4.1	& 1.1	& --0.47 & 0.04	& 35 & 0.05 & 7 & 0.45	& 0.09	&	0.04 &	3 & 0.24 & 0.03	& 4 \\
 HD 18757	& 5600	& 4.1	& 0.8	& --0.25 & 0.04	& 34 & 0.01 & 7 & 0.25	& 0.03	&	0.02 &	5 & 0.24 & 0.02	& 3 \\
 HD 21543	& 5640	& 4.1	& 1.0	& --0.53 & 0.04	& 34 & 0.05 & 7 & 0.45	& 0.05	&	0.02 &	3 & 0.38 & 0.07	& 4 \\
 HD 24156	& 5470	& 3.9	& 0.9	& --0.43 & 0.05	& 38 & 0.02 & 7 & 0.51	& 0.07	&	0.01 &	4 & 0.27 & 0.05	& 4 \\
 HD 29587	& 5660	& 4.2	& 0.9	& --0.51 & 0.04	& 34 & 0.06 & 7 & ...   & --0.04 &	0.02 &	3 & 0.26 & 0.03	& 3 \\
 HD 30649	& 5770	& 4.1	& 0.9	& --0.45 & 0.04	& 36 & 0.02 & 7 & 0.54	& 0.01	&	0.03 &	4 & 0.29 & 0.05	& 4 \\
 HD 37739	& 6280	& 3.8	& 1.0	& --0.41 & 0.05	& 24 & 0.05 & 5 & 0.40	& 0.10	&	0.01 &	3 & 0.33 & 0.12	& 4 \\
 HD 38767	& 6170	& 3.6	& 1.0	& --0.55 & 0.05	& 26 & 0.06 & 5 & ...   & 0.12	&	0.04 &	3 & 0.22 & 0.05	& 3 \\
 HD 96094	& 5970	& 4.1	& 1.1	& --0.31 & 0.05	& 26 & 0.04 & 6 & 0.25	& --0.09 &	0.04 &	3 & 0.08 & 0.03	& 3 \\
 HD 114606	& 5600	& 4.2	& 0.9	& --0.50 & 0.05	& 31 & 0.05 & 7 & ...  	& 0.06	&	0.03 &	3 & 0.34 & 0.05	& 4 \\
 HD 121533	& 5600	& 4.0	& 1.0	& --0.37 & 0.05	& 28 & 0.04 & 6 & ...   & 0.07	&	0.04 &	5 & 0.33 & 0.05	& 4 \\
 HD 131582	& 4770	& 4.3	& 0.8	& --0.36 & 0.03	& 37 & 0.02 & 5 & ...   & 0.07	&	0.05 &	4 & 0.17 & 0.03	& 4 \\
 HD 132142	& 5180	& 4.3	& 0.8	& --0.35 & 0.03	& 34 & 0.05 & 7 & 0.22	& 0.09	&	0.06 &	4 & 0.31 & 0.02	& 4 \\
 HD 133621	& 5650	& 3.7	& 1.0	& --0.40 & 0.05	& 34 & 0.06 & 7 & ...   & 0.11	&	0.03 &	3 & 0.31 & 0.03	& 4 \\
 HD 137687	& 5070	& 3.6	& 0.8	& --0.56 & 0.03	& 29 & 0.02 & 6 & 0.48	& 0.09	&	0.04 &	2 & 0.47 & 0.10	& 4 \\
 HD 139457	& 6000	& 3.9	& 1.2	& --0.43 & 0.06	& 24 & 0.04 & 7 & 0.50	& 0.03	&	0.03 &	3 & 0.16 & 0.06	& 3 \\
 HD 143291	& 5280	& 4.4	& 0.7	& --0.33 & 0.03	& 36 & 0.05 & 6 & ...   & --0.03 &	0.05 &	3 & 0.18 & 0.04	& 4 \\
 HD 152123	& 6040	& 3.7	& 1.2	& --0.16 & 0.05	& 23 & 0.05 & 6 & ...   & 0.06	&	0.04 &	3 & 0.24 & 0.05	& 4 \\
 HD 156802	& 5660	& 3.9	& 0.9	& --0.37 & 0.05	& 32 & 0.04 & 7 & ...   & 0.10	&	0.01 &	3 & 0.26 & 0.04	& 3 \\
 HD 158226	& 5740	& 4.0	& 1.1	& --0.47 & 0.05	& 30 & 0.03 & 7 & 0.42	& 0.06	&	0.03 &	3 & 0.39 & 0.03	& 4 \\
 HD 165401	& 5770	& 4.3	& 0.8	& --0.33 & 0.05	& 34 & 0.05 & 7 & ...   & 0.06	&	0.03 &	3 & 0.24 & 0.01	& 3 \\
 HD 170357	& 5710	& 3.9	& 0.9	& --0.34 & 0.04	& 31 & 0.06 & 6 & 0.30	& 0.02	&	0.01 &	3 & 0.30 & 0.02	& 4 \\
 HD 190404	& 5000	& 4.5	& 0.8	& --0.58 & 0.04	& 35 & 0.03 & 5 & ...   & 0.03	&	0.02 &	3 & 0.25 & 0.03	& 3 \\
 HD 200580	& 5860	& 3.9	& 1.0	& --0.56 & 0.05	& 28 & 0.03 & 7 & 0.21	& 0.12	&	0.03 &	3 & 0.14 & 0.02	& 4 \\
 HD 201099	& 5890	& 3.8	& 1.0	& --0.40 & 0.05	& 31 & 0.04 & 7 & ...   & --0.04 &	0.04 &	3 & 0.16 & 0.07	& 4 \\
 HD 215594	& 5810	& 3.8	& 1.0	& --0.26 & 0.04	& 33 & 0.03 & 7 & ...   & --0.03 &	0.02 &	3 & 0.12 & 0.04	& 4 \\
 HD 221830	& 5720	& 4.1	& 1.0	& --0.37 & 0.05	& 32 & 0.05 & 7 & 0.39	& 0.03	&	0.02 &	3 & 0.31 & 0.03	& 3 \\
 HD 224817	& 5720	& 3.7	& 0.8	& --0.60 & 0.04	& 31 & 0.04 & 5 & 0.25	& 0.06	&	0.02 &	3 & 0.23 & 0.04	& 3 \\
\noalign{\smallskip}
\hline
\noalign{\smallskip}
HD 41330 & 5820	& 4.0	& 1.0	& --0.16 & 0.04	& 35 & 0.03 &	7 & 0.19  & --0.06 & 0.03 & 3 &	0.09	& 0.05	& 4 \\
HD 43318 & 6230	& 3.6	& 1.1	& --0.10 & 0.05	& 29 & 0.05 &	6 & 0.00  & --0.02 & 0.05 & 3 &	0.07	& 0.33	& 4 \\
HD 69897 & 6330	& 4.0	& 1.3	& --0.24 & 0.05	& 38 & 0.02 &	7 & --0.01 & --0.01 & 0.02 & 3 & 0.19	& 0.09	& 4 \\
HD 108954 & 5960 & 3.9	& 1.1	& --0.07 & 0.05	& 29 & 0.03 &	6 & --0.01 & 0.01 & 0.02	& 3 &	0.11	& 0.02	& 4 \\
HD 153597 & 6380 & 4.0	& 1.3	& --0.10 & 0.05	& 20 & 0.05 &	5 & 0.15  & 0.02 & 0.02	& 2 &	0.19	& 0.33	& 4 \\
HD 157466 & 6130 & 4.2	& 1.2	& --0.34 & 0.05	& 30 & 0.05 &	7 & 0.24  & --0.07 & 0.04 & 3 &	--0.02	& 0.03	& 3 \\
HD 176377 & 5770 & 4.1	& 0.8	& --0.24 & 0.05	& 37 & 0.05 &	7 & ...   & --0.06 & 0.01 & 3 &	0.02	& 0.01	& 4 \\
\noalign{\smallskip}
\hline
\hline
Star & [Al/Fe] & $\sigma$ & n & [Si/Fe] &$\sigma$ &  n& [Ca/Fe]&$\sigma$ &  n& [Sc/Fe]& $\sigma$& n& [Ti{\sc i}/Fe]& $\sigma$& n\\
\hline
 BD +68 813  & 0.32 & 0.03 & 2 &	0.23	& 0.04	& 14 &	0.24 &	0.05 &	8  & 0.20 & 0.03 & 12	& 0.24	& 0.04	& 10	\\                
BD +31 3330 & 0.10 & 0.05 & 3 &	0.06	& 0.05	& 18 &	0.05 &	0.05 &	10 & 0.03 & 0.05 & 9	& 0.09	& 0.05	& 23	\\                
HD 10519    & 0.32 & 0.04 & 2 &	0.24	& 0.05	& 19 &	0.27 &	0.05 &	6  & 0.22 & 0.05 & 10	& 0.28	& 0.05	& 10	\\                
HD 12782    & 0.38 & 0.03 & 2 &	0.26	& 0.05	& 16 &	0.27 &	0.06 &	11 & 0.17 & 0.06 & 8	& 0.29	& 0.05	& 23	\\                
HD 16397    & 0.25 & 0.03 & 2 &	0.16	& 0.05	& 15 &	0.17 &	0.05 &	8  & 0.10 & 0.05 & 11	& 0.19	& 0.04	& 10	\\                
HD 18757    & 0.27 & 0.00 & 2 &	0.18	& 0.05	& 20 &	0.15 &	0.05 &	10 & 0.15 & 0.05 & 10	& 0.20	& 0.05	& 22	\\                
HD 21543    & 0.29 & 0.01 & 2 &	0.20	& 0.05	& 16 &	0.22 &	0.05 &	9  & 0.15 & 0.04 & 11	& 0.27	& 0.05	& 14	\\                
HD 24156    & 0.28 & 0.04 & 2 &	0.21	& 0.06	& 16 &	0.20 &	0.05 &	8  & 0.20 & 0.04 & 12	& 0.21	& 0.05	& 19	\\                
HD 29587    & 0.23 & 0.00 & 2 &	0.16	& 0.05	& 16 &	0.18 &	0.05 &	7  & 0.10 & 0.05 & 10	& 0.24	& 0.05	& 14	\\                
HD 30649    & 0.19 & 0.04 & 2 &	0.17	& 0.05	& 20 &	0.20 &	0.05 &	9  & 0.15 & 0.05 & 11	& 0.21	& 0.05	& 10	\\                
HD 37739    & 0.23 & ... & 1 &	0.27	& 0.06	& 12 &	0.28 &	0.04 &	5  & 0.20 & 0.06 & 10	& 0.32	& 0.05	& 6	\\                
HD 38767    & ...  & ...  & ... &	0.14	& 0.05	& 14 &	0.18 &	0.04 &	6  & 0.08 & 0.05 & 8	& 0.25	& 0.05	& 5	\\        
HD 96094    & 0.13 & ... & 1 &	0.02	& 0.05	& 12 &	0.05 &	0.05 &	6  & 0.15 & 0.05 & 9	& 0.09	& 0.05	& 9	\\                
HD 114606   & 0.23 & 0.01 & 2 &	0.25	& 0.05	& 15 &	0.26 &	0.05 &	8  & 0.22 & 0.04 & 10	& 0.30	& 0.05	& 14	\\                
HD 121533   & 0.33 & 0.05 & 3 &	0.20	& 0.05	& 14 &	0.20 &	0.05 &	7  & 0.11 & 0.04 & 10	& 0.22	& 0.04	& 10	\\                
HD 131582   & 0.23 & 0.04 & 2 &	0.08	& 0.05	& 17 &	0.14 &	0.05 &	5  & 0.10 & 0.05 & 9	& 0.18	& 0.05	& 21	\\                
HD 132142   & 0.34 & 0.02 & 2 &	0.17	& 0.05	& 15 &	0.26 &	0.05 &	10 & 0.14 & 0.05 & 11	& 0.30	& 0.05	& 23	\\                
HD 133621   & 0.32 & 0.01 & 2 &	0.24	& 0.05	& 18 &	0.20 &	0.05 &	8  & 0.12 & 0.04 & 10	& 0.23	& 0.04	& 9	\\                
HD 137687   & 0.52 & 0.05 & 2 &	0.31	& 0.05	& 16 &	0.29 &	0.05 &	11 & 0.23 & 0.05 & 9	& 0.33	& 0.05	& 15	\\                
HD 139457   & 0.10 & ... & 1 &	0.09	& 0.05	& 13 &	0.12 &	0.05 &	7  & 0.07 & 0.03 & 8	& 0.14	& 0.05	& 6	\\                
HD 143291   & 0.17 & 0.03 & 2 &	0.06	& 0.04	& 16 &	0.13 &	0.05 &	10 & 0.06 & 0.03 & 11	& 0.16	& 0.05	& 23	\\                
HD 152123   & 0.25 & 0.05 & 2 &	0.09	& 0.05	& 12 &	0.20 &	0.03 &	6  & 0.18 & 0.04 & 9	& 0.21	& 0.05	& 11	\\                
HD 156802   & 0.24 & 0.05 & 2 &	0.18	& 0.05	& 16 &	0.22 &	0.05 &	7  & 0.13 & 0.04 & 9	& 0.22	& 0.05	& 10	\\                
HD 158226   & 0.36 & 0.03 & 3 &	0.23	& 0.04	& 15 &	0.27 &	0.06 &	8  & 0.21 & 0.04 & 11	& 0.28	& 0.05	& 11	\\                
HD 165401   & 0.18 & 0.04 & 2 &	0.16	& 0.04	& 18 &	0.14 &	0.05 &	9  & 0.13 & 0.04 & 10	& 0.18	& 0.03	& 11	\\                
HD 170357   & 0.25 & 0.01 & 2 &	0.15	& 0.05	& 17 &	0.18 &	0.05 &	9  & 0.18 & 0.04 & 11	& 0.22	& 0.05	& 12	\\                
HD 190404   & 0.32 & 0.06 & 4 &	0.21	& 0.05	& 17 &	0.28 &	0.06 &	9  & 0.22 & 0.05 & 9	& 0.31	& 0.05	& 23	\\                
HD 200580   & 0.22 & 0.03 & 2 &	0.09	& 0.05	& 14 &	0.15 &	0.05 &	6  & --0.05	& 0.04	& 9	& 0.15	& 0.05	& 8	\\        
HD 201099   & ...  & ...  & ... &	0.05	& 0.05	& 15 &	0.09 &	0.05 &	10 & --0.01	& 0.04	& 10	& 0.09	& 0.04	& 9	\\
HD 215594   & 0.14 & 0.02 & 2 &	0.09	& 0.05	& 19 &	0.11 &	0.05 &	9  & 0.03 & 0.05 & 9	& 0.07	& 0.05	& 11	\\                
HD 221830   & 0.24 & ... & 1 &	0.23	& 0.05	& 17 &	0.26 &	0.05 &	8  & 0.16 & 0.04 & 5	& 0.32	& 0.04	& 15	\\                
HD 224817   & 0.21 & 0.01 & 2 &	0.20	& 0.05	& 16 &	0.19 &	0.05 &	8  & 0.15 & 0.05 & 9	& 0.18	& 0.04	& 8	\\                
\noalign{\smallskip}
\hline
\noalign{\smallskip}
HD 41330    & 0.11	& 0.05	& 2 &	0.03	& 0.04	& 20 &	0.06 &	0.05 &	10 & 0.05	& 0.05	& 11	& 0.04	& 0.05	& 14	\\
HD 43318    & --0.06	& 0.02	& 3 &	0.05	& 0.05	& 5  &	0.07 &	0.05 &	8  & 0.01	& 0.04	& 10	& 0.02	& 0.05	& 8	\\
HD 69897    & --0.04	& 0.04	& 2 &	0.08	& 0.04	& 16 &	0.11 &	0.04 &	8  & --0.04	& 0.05	& 9	& 0.09	& 0.05	& 8	\\
HD 108954   & 0.02	& 0.04	& 3 &	0.02	& 0.05	& 15 &	0.02 &	0.05 &	8  & 0.00	& 0.05	& 6	& 0.04	& 0.05	& 9	\\
HD 153597   & ...  	& ...  	& ... &	0.08	& 0.05	& 10 &	0.08 &	0.04 &	5  & --0.03	& 0.05	& 7	& 0.06	& 0.06	& 5	\\
HD 157466   & ...  	& ...  	& ... &	--0.02	& 0.04	& 15 &	0.06 &	0.05 &	7  & 0.04 & 0.05 & 9	& --0.03	& 0.04	& 7	\\
HD 176377   & 0.05	& 0.02	& 3 &	0.01	& 0.05	& 16 &	0.06 &	0.05 &	8  & --0.04	& 0.04	& 10	& 0.01 & 0.04	& 16	\\
\noalign{\smallskip}
\hline
\hline
Star & [Ti{\sc ii}/Fe] & $\sigma$ & n & [V/Fe] & $\sigma$ & n & [Cr/Fe] &$\sigma$ &  n& [Co/Fe]& $\sigma$& n& [Ni/Fe]& $\sigma$& n\\
\hline
BD +68 813  & 0.21  & 0.03 & 3 & 0.07	& 0.04	& 6	& 0.03	& 0.05	& 13	& 0.09	& 0.03	& 3	& 0.02	& 0.05	& 19 \\
BD +31 3330 & 0.06  & 0.03 & 3 & 0.07	& 0.05	& 11	& 0.00	& 0.05	& 18	& --0.02 & 0.05	& 8	& -0.02	& 0.05	& 27 \\
HD 10519    & 0.31  & 0.03 & 3 & 0.17	& 0.04	& 7	& 0.03	& 0.05	& 13	& 0.13	& 0.04	& 8	& 0.01	& 0.05	& 23 \\
HD 12782    & 0.29  & 0.05 & 3 & 0.11	& 0.05	& 11	& 0.06	& 0.06	& 16	& 0.14	& 0.03	& 9	& 0.05	& 0.05	& 25 \\
HD 16397    & 0.16  & 0.05 & 3 & 0.10	& 0.05	& 9	& --0.02 & 0.04	& 15	& 0.05	& 0.04	& 6	& -0.02	& 0.05	& 23 \\
HD 18757    & 0.15  & 0.05 & 3 & 0.06	& 0.05	& 14	& 0.03	& 0.06	& 19	& 0.07	& 0.05	& 13	& -0.01	& 0.05	& 29 \\
HD 21543    & 0.23  & 0.05 & 3 & 0.10	& 0.05	& 11	& 0.05	& 0.05	& 16	& 0.12	& 0.05	& 5	& 0.02	& 0.03	& 23 \\
HD 24156    & 0.25  & 0.03 & 3 & 0.09	& 0.05	& 13	& 0.04	& 0.06	& 15	& 0.10	& 0.05	& 8	& 0.04	& 0.05	& 24 \\
HD 29587    & 0.18  & 0.05 & 3 & 0.14	& 0.05	& 13	& 0.03	& 0.05	& 16	& 0.10	& 0.05	& 8	& 0.03	& 0.05	& 27 \\
HD 30649    & 0.22  & 0.03 & 3 & 0.02	& 0.05	& 10	& 0.02	& 0.05	& 15	& 0.09	& 0.05	& 6	& 0.01	& 0.05	& 28 \\
HD 37739    & 0.33  & 0.02 & 2 & ...   	& ...   & ...  	& 0.02	& 0.05	& 11	& 0.02	& 0.02	& 3	& 0.08	& 0.06	& 13 \\
HD 38767    & 0.19  & 0.01 & 2 & ...   	& ...   & ...  	& 0.04	& 0.04	& 9	& ...  	& ...  	& ...  	& -0.05	& 0.06	& 12 \\
HD 96094    & 0.14  & 0.04 & 3 & 0.12	& 0.04	& 4	& --0.03 & 0.05	& 13	& --0.01 & 0.03	& 4	& -0.04	& 0.05	& 22 \\
HD 114606   & 0.23  & 0.06 & 3 & 0.14	& 0.03	& 6	& 0.05	& 0.06	& 15	& 0.14	& 0.02	& 4	& 0.05	& 0.05	& 22 \\
HD 121533   & 0.17  & 0.03 & 3 & 0.01	& 0.05	& 11	& 0.01	& 0.05	& 12	& 0.09	& 0.04	& 6	& 0.00	& 0.05	& 15 \\
HD 131582   & 0.07  & 0.03 & 3 & 0.18	& 0.05	& 3	& 0.08	& 0.04	& 14	& 0.06	& 0.05	& 8	& 0.03	& 0.05	& 21 \\
HD 132142   & 0.25  & 0.05 & 3 & 0.18	& 0.04	& 8	& 0.07	& 0.04	& 16	& 0.14	& 0.04	& 7	& 0.03	& 0.05	& 24 \\
HD 133621   & 0.19  & 0.05 & 3 & 0.12	& 0.05	& 7	& 0.05	& 0.05	& 13	& 0.06	& 0.05	& 6	& 0.02	& 0.05	& 21 \\
HD 137687   & 0.30  & 0.05 & 3 & 0.17	& 0.03	& 9	& 0.10	& 0.04	& 13	& 0.16	& 0.05	& 7	& 0.06	& 0.05	& 22 \\
HD 139457   & 0.16  & 0.05 & 3 & 0.05	& 0.06	& 3	& 0.01	& 0.05	& 12	& 0.03	& 0.04	& 3	& -0.05	& 0.05	& 15 \\
HD 143291   & 0.10  & 0.02 & 3 & 0.09	& 0.05	& 13	& 0.06	& 0.06	& 19	& 0.01	& 0.05	& 8	& -0.03	& 0.05	& 26 \\
HD 152123   & 0.14  & 0.03 & 2 & ...   	& ...  	& ...  	& -0.01	& 0.05	& 14	& 0.08	& 0.05	& 5	& 0.00	& 0.05	& 17 \\
HD 156802   & 0.13  & 0.05 & 3 & 0.10	& 0.05	& 7	& 0.01	& 0.06	& 12	& 0.15	& 0.05	& 5	& 0.02	& 0.05	& 25 \\
HD 158226   & 0.24  & 0.06 & 3 & 0.16	& 0.05	& 7	& 0.01	& 0.06	& 13	& 0.12	& 0.04	& 6	& 0.03	& 0.05	& 23 \\
HD 165401   & 0.18  & 0.05 & 3 & 0.05	& 0.05	& 8	& --0.03 & 0.05	& 13	& 0.09	& 0.05	& 6	& --0.01 & 0.03	& 24 \\
HD 170357   & 0.20  & 0.02 & 3 & 0.08	& 0.05	& 10	& 0.01	& 0.07	& 15	& 0.11	& 0.05	& 7	& 0.01	& 0.05	& 27 \\
HD 190404   & 0.24  & 0.05 & 3 & 0.19	& 0.05	& 9	& 0.07	& 0.04	& 12	& 0.10	& 0.05	& 8	& 0.04	& 0.05	& 25 \\
HD 200580   & 0.06  & 0.05 & 3 & 0.15	& 0.05	& 6	& 0.00	& 0.05	& 10	& --0.01 & 0.02	& 3	& --0.04 & 0.05	& 15 \\
HD 201099   & 0.05  & 0.03 & 2 & 0.00	& 0.05	& 7	& 0.00	& 0.05	& 17	& --0.07 & 0.03	& 4	& --0.03 & 0.05	& 17 \\
HD 215594   & 0.10  & 0.02 & 2 & 0.02	& 0.05	& 8	& --0.03 & 0.05	& 14	& --0.02 & 0.04	& 7	& -0.08	& 0.05	& 25 \\
HD 221830   & 0.28  & 0.04 & 3 & 0.08	& 0.04	& 4	& 0.03	& 0.06	& 6	& 0.08	& 0.05	& 6	& 0.01	& 0.05	& 22 \\
HD 224817   & 0.21  & 0.04 & 3 & 0.04	& 0.05	& 7	& 0.01	& 0.05	& 11	& 0.07	& 0.04	& 5	& -0.01	& 0.05	& 20 \\
\noalign{\smallskip}
\hline
\noalign{\smallskip}
HD 41330  &   0.03  & 0.00	& 2 &	--0.02	& 0.04	& 9	& --0.01 & 0.06	& 18 &	--0.01	& 0.05	& 9 &	--0.04	& 0.05	& 28 \\
HD 43318  &   0.01  & 0.00	& 3 &	0.00	& 0.04	& 3	& 0.00	&  0.05	& 11 &	--0.04	& 0.01	& 4 &	--0.05	& 0.04	& 14 \\
HD 69897  &   0.08  & 0.01	& 2 &	--0.06	& 0.03	& 2	& 0.01	&  0.05	& 13 &	--0.05	& 0.05	& 3 &	--0.06	& 0.05	& 21 \\
HD 108954 &  0.02  & ...	& 1 &	...   	& ...  	& ...  	& 0.04	&  0.05	& 15 &	--0.01	& 0.04	& 4 &	--0.03	& 0.05	& 17 \\
HD 153597 &  0.06  & ...	& 1 &	0.09	& 0.01	& 3	& 0.06	&  0.05	& 10 &	0.08	& 0.04	& 8 &	--0.06	& 0.05	& 11 \\
HD 157466 &  --0.04 & 0.01	& 2 &	0.04	& 0.04	& 5	& --0.02 & 0.05	& 14 &	--0.04	& 0.04	& 3 &	--0.04	& 0.05	& 21 \\
HD 176377 &  0.00  & 0.01	& 3 &	--0.05	& 0.05	& 8	& 0.03	&  0.05	& 15 &	--0.04	& 0.04	& 7 &	--0.04	& 0.05	& 27 \\
\hline
\end{longtable}
}

\longtab{5}{
\begin{longtable}{lccccccccccccccc}
\caption{Elemental abundances of neutron-capture elements for the programme and comparison stars.} \\
\hline\hline 
Star & [Y/Fe] & $\sigma$ & n & [Zr{\sc i}/Fe] & $\sigma$ & n & [Zr{\sc ii}/Fe] & $\sigma$ & n &[Ba/Fe] & [La/Fe]& $\sigma$& n \\
\hline
\label{table:5}
\endfirsthead
\caption{continued.}\\
\endhead
BD +68 813	& --0.14 & 0.08	& 4 &	...   & ...  	& ... &  ...	&...	&...  & --0.17	& 0.09		&...	&1 \\ 
BD +31 3330	& 0.05	 & 0.01	& 3 &	0.09  & 0.04	& 7   & ...	&...	&...  & --0.08	& 0.11		&0.01	&3 \\ 
HD 10519	& --0.05 & 0.05	& 6 &	0.22  & 0.03	& 3   & 0.08	&0.07	&2    & --0.12	& 0.08		&0.06	&4 \\ 
HD 12782	& --0.01 & 0.05	& 5 &	0.04  & 0.05	& 7   & 0.05	&0.08	&2    & ...	& 0.13		&0.06	&4 \\ 
HD 16397	& --0.14 & 0.07	& 6 &	0.12  & 0.07	& 3   & 0.09	&...	&1    & ...	& --0.05	&0.04	&2 \\ 
HD 18757	& --0.16 & 0.07	& 4 &	--0.03& 0.04	& 3   & --0.13	&0.02	&2    & --0.10	& --0.10	&0.03	&4 \\ 
HD 21543	& --0.07 & 0.05	& 4 &	0.04  & 0.06	& 4   & ...	&...	&...  & --0.17	& --0.06	&...	&1 \\ 
HD 24156	& 0.00	 & 0.07	& 7 &	--0.07& 0.04	& 2   & --0.02	&...	&1    & --0.08	& --0.08	&0.02	&2 \\ 
HD 29587	& --0.11 & 0.06	& 7 &	0.07  & 0.07	& 4   & 0.07	&0.01	&2    & --0.12	& --0.04	&...	&1 \\ 
HD 30649	& --0.06 & 0.11	& 6 &	0.11  & 0.05	& 5   & 0.05	&0.00	&2    & --0.05	& --0.01	&0.04	&3 \\ 
HD 37739	& --0.17 & 0.04	& 4 &	...   & ...  	& ... &  0.12	&...	&1    & --0.02	& 0.08		&...	&1 \\ 
HD 38767	& 0.01   & 0.04	& 4 &	...   & ...  	& ... &  0.22	&...	&1    & ...	& 0.08		&0.11	&2 \\ 
HD 96094	& --0.14 & 0.05	& 5 &	0.16  & 0.00   	& 2   & 0.05	&...	&1    & --0.10	& 0.09		&0.01	&2 \\ 
HD 114606	& --0.12 & 0.04	& 5 &	0.28  & 0.05	& 3   & 0.20	&...	&1    & --0.20	& 0.10		&0.06	&2 \\ 
HD 121533	& 0.02	 & 0.05	& 2 &	0.01  & ...   	& 1   & --0.04	&0.08	&2    & --0.08	& 0.03		&0.01	&2 \\ 
HD 131582	& --0.16 & 0.06	& 2 &	0.06  & 0.10 	& 8   & 0.10	&...	&1    & --0.16	& 0.00		&...	&1 \\ 
HD 132142	& --0.11 & 0.04	& 2 &	0.03  & 0.07	& 6   & ...	&...	&...  & 0.02	& ...		&...	&...\\
HD 133621	& --0.20 & 0.04	& 2 &	--0.05& ...   	& 1   & 0.01	&...	&1    & --0.10	& --0.04	&0.10	&2 \\ 
HD 137687	& 0.05	 & 0.06	& 5 &	0.14  & 0.08	& 6   & 0.23	&0.04	&2    & --0.14	& 0.02		&0.06	&2 \\ 
HD 139457	& --0.01 & 0.03	& 6 &	...   & ...  	& ... &  0.02	&...	&1    & ...	& 0.07		&0.02	&3 \\ 
HD 143291	& --0.16 & 0.05	& 5 &	0.06  & 0.05	& 6   & 0.06	&...	&1    & --0.17	& 0.09		&0.01	&2 \\ 
HD 152123	& 0.04	 & 0.00	& 2 &	...   & ...  	& ... &  ...	&...	&...  & ...	& 0.08		&...	&1 \\ 
HD 156802	& --0.08 & 0.05	& 6 &	--0.09& 0.01	& 2   & --0.10	&...	&1    & --0.13	& 0.03		&0.03	&2 \\ 
HD 158226	& --0.13 & 0.06	& 4 &	0.21  & 0.08	& 2   & 0.13	&0.03	&2    & --0.21	& 0.06		&0.02	&3 \\ 
HD 165401	& --0.17 & 0.01	& 2 &	--0.02& 0.03	& 3   & --0.07	&...	&1    & --0.07	& --0.07	&...	&1 \\ 
HD 170357	& --0.14 & 0.06	& 6 &	--0.02& 0.04	& 3   & 0.00	&...	&1    & --0.06	& 0.04		&0.04	&3 \\ 
HD 190404	& 0.07	 & 0.11	& 4 &	0.18  & 0.07	& 7   & 0.15	&...	&1    & ...	& 0.02		&0.04	&2 \\ 
HD 200580	& --0.16 & 0.05	& 4 &	...   & ...  	& ... &  0.11	&...	&1    & --0.08	& --0.02	&...	&1 \\ 
HD 201099	& --0.21 & 0.06	& 4 &	0.21  & 0.01	& 3   & 0.09	&...	&1    & 0.00	& --0.05	&...	&1 \\ 
HD 215594	& --0.14 & 0.09	& 4 &	0.06  & 0.09	& 3   & 0.06	&...	&1    & --0.04	& --0.01	&...	&1 \\ 
HD 221830	& 0.05	 & 0.06	& 5 &	0.06  & 0.12	& 2   & 0.14	&0.04	&2    & --0.10	& 0.12		&0.06	&2 \\ 
HD 224817	& --0.16 & 0.06	& 7 &	0.22  & 0.06	& 2   & ...	&...	&...  & --0.10	& --0.08	&0.07	&3 \\ 
\noalign{\smallskip}
\hline
\noalign{\smallskip}
HD 41330	 &--0.11	&0.07	&5	&--0.01	&0.05	&2	&--0.09	&0.03	&2	&0.01	&--0.09	&0.05	&3 \\   
HD 43318	 &0.03		&0.07	&4	&...	&...	&...	&0.07	&0.00	&2	&...	&0.10	&0.09	&3 \\   
HD 69897	 &--0.06	&0.04	&6	&0.10	&0.01	&2	&0.12	&0.00	&2	&0.04	&0.02	&0.05	&2 \\   
HD 108954 &--0.03	&0.04	&2	&0.05	&...	&1	&--0.07	&...	&1	&--0.09	&0.07	&0.08	&3 \\   
HD 153597 &--0.01	&0.04	&3	&...	&...	&...	&...	&...	&...	&0.07	&...	&...	&... \\ 
HD 157466 &--0.10	&0.08	&4	&0.17	&0.05	&3	&0.15	&0.04	&2	&--0.04	&0.06	&0.01	&2 \\   
HD 176377 &--0.06	&0.04	&6	&0.11	&0.06	&5	&0.07	&0.00	&2	&0.04	&0.10	&0.04	&2 \\   
\noalign{\smallskip}  
\hline\hline 
Star & [Ce/Fe] & $\sigma$ & n & [Pr/Fe] & $\sigma$ & n & [Nd/Fe] &  $\sigma$ & n &  [Sm/Fe]  &  $\sigma$ & n &  [Eu/Fe]  &  $\sigma$ & n \\
\hline
BD +68 813  &--0.04	&0.00	&2	&0.41	&...	&1	&0.22	&0.10	&4	&0.48	&0.04	&2  &0.54	&0.14	&2 \\
BD +31 3330 &0.03	&...	&1	&0.35	&0.06	&2	&0.18	&0.09	&3	&0.34	&0.07	&2  &0.39	&0.16	&2 \\
HD 10519   &0.00	&0.10	&4	&0.29	&0.08	&2	&0.02	&0.02	&5	&0.22	&0.04	&3  &0.24	&0.01	&2 \\
HD 12782   &--0.11	&0.06	&4	&0.29	&...	&1	&0.24	&0.05	&7	&0.34	&0.06	&2  &0.43	&0.04	&2 \\ 
HD 16397   &--0.07	&0.05	&2	&0.21	&0.03	&2	&0.02	&0.06	&5	&0.18	&0.04	&3  &0.25	&0.12	&2 \\
HD 18757   &--0.05	&0.03	&3	&0.17	&...	&1	&0.01	&0.07	&7	&0.09	&0.08	&3  &0.15	&0.01	&2 \\
HD 21543   &0.00	&...	&1	&0.27	&0.05	&2	&0.05	&0.05	&4	&0.31	&0.06	&3  &0.34	&0.08	&2 \\
HD 24156   &0.06	&0.02	&2	&0.25	&...	&1	&0.11	&0.03	&6	&0.21	&0.02	&3  &0.34	&0.12	&2 \\
HD 29587   &0.06	&0.04	&4	&0.41	&...	&1	&0.16	&0.02	&4	&0.30	&0.00	&3  &0.39	&0.09	&2 \\
HD 30649   &0.11	&0.12	&5	&0.38	&...	&1	&0.09	&0.01	&6	&0.34	&0.04	&3  &0.37	&0.12	&2 \\
HD 37739   &...		&...	&... &...	&...	&...&--0.02	&...	&1	&0.10	&...	&1  &0.11	&...	&1 \\
HD 38767   &0.09	&...	&1	&0.28	&...	&1	&0.10	&0.01	&4	&0.25	&...	&1  &0.17	&...	&1 \\
HD 96094   &0.11	&0.01	&2	&0.18	&...	&1	&0.07	&0.04	&3	&0.18	&0.07	&2  &0.22	&0.04	&2 \\
HD 114606  &0.06	&0.02	&3	&0.30	&...	&1	&0.13	&0.01	&4	&0.39	&0.04	&2  &0.32	&...	&1 \\
HD 121533  &0.02	&0.07	&3	&0.32	&...	&1	&0.09	&0.07	&5	&0.32	&...	&1  &0.29	&0.11	&2 \\
HD 131582  &0.02	&0.08	&2	&...	&...	&...	&0.14	&0.01	&2	&0.33	&0.05	&2  &0.26	&0.01	&2 \\
HD 132142  &--0.11	&0.01	&3	&0.26	&0.03	&2	&0.02	&0.05	&4	&0.19	&0.05	&3  &0.26	&0.04	&2 \\
HD 133621  &--0.07	&0.09	&3	&0.18	&...	&1	&--0.07	&0.03	&5	&0.15	&0.07	&3  &0.12	&...	&2 \\
HD 137687  &--0.02	&0.07	&3	&0.28	&0.06	&2	&0.19	&0.05	&6	&0.36	&...	&1  &0.34	&0.04	&2 \\
HD 139457  &--0.07	&...	&1	&0.35	&...	&1	&0.15	&0.03	&3	&0.17	&0.06	&3  &0.23	&0.19	&2 \\
HD 143291  &0.00	&0.04	&4	&0.29	&...	&1	&0.07	&0.05	&4	&0.19	&0.08	&3  &0.25	&0.11	&2 \\
HD 152123  & ...		&...	&...&...	&...	&...&0.04	&0.01	&2	&0.32	&...	&1  &0.21	&...	&1 \\
HD 156802  &--0.08	&0.04	&3	&0.32	&...	&1	&0.16	&0.06	&4	&0.32	&...	&1  &0.27	&0.05	&2 \\
HD 158226  &--0.02	&0.05	&3	&0.23	&...	&1	&0.10	&0.05	&4	&0.36	&0.06	&3  &0.31	&0.06	&2 \\
HD 165401  &--0.01	&...	&1	&...	&...	&...&0.07	&0.07	&5	&0.26	&0.03	&2  &0.27	&0.02	&2 \\
HD 170357  &0.04	&0.03	&3	&0.13	&0.01	&2	&--0.01	&0.06	&5	&0.23	&0.06	&3  &0.23	&0.00	&2 \\
HD 190404  &0.05	&0.07	&2	&...	&...	&...&0.20	&0.03	&4	&0.34	&0.10	&3  &0.41	&0.04	&2 \\
HD 200580  &--0.08	&...	&1	&0.11	&0.03	&2	&0.03	&0.03	&4	&0.12	&0.08	&2  &0.02	&...	&1 \\
HD 201099  &--0.11	&0.03	&2	&0.35	&...	&1	&0.00	&0.04	&3	&0.12	&...	&1  &0.18	&0.20	&2 \\
HD 215594  &--0.08	&0.08	&3	&...	&...	&...&--0.05	&0.07	&5	&0.06	&0.07	&3  &0.13	&0.18	&2 \\
HD 221830  &--0.03	&0.02	&3	&0.23	&0.07	&2	&0.13	&0.05	&3	&0.29	&0.04	&2  &0.46	&0.01	&2 \\
HD 224817  &0.00	&0.02	&3	&0.14	&0.08	&2	&0.01	&0.08	&5	&0.16	&0.07	&2  &0.20	&0.04	&2 \\
\noalign{\smallskip}
\hline
\noalign{\smallskip}			  		  		  		  		  				
HD 41330    &0.02		&0.04	&2	&0.12	&0.06	&2	&0.04	&0.02	&5	&0.07	&...	&1  &0.04	&0.06	&2	\\
HD 43318    &--0.06		&0.06	&3	&0.16	&...	&1	&--0.03	&0.05	&3	&0.10	&...	&1  &0.07	&0.00	&2	\\
HD 69897    &0.04		&0.03	&3	&0.20	&...	&1	&0.04	&0.09	&5	&0.13	&0.04	&2  &0.14	&0.08	&2	\\
HD 108954   &0.03		&0.06	&2	&0.03	&...	&1	&--0.02	&0.03	&3	&0.15	&...	&1  &...	&...	&...	\\
HD 153597   &...			&...	&...	&0.22	&...	&1	&0.04	&0.00	&2	&...	&...	&...&0.05	&...	&1	\\
HD 157466   &0.02		&0.04	&2	&0.30	&...	&1	&0.02	&0.00	&2	&...	&...	&...&0.11	&...	&1	\\
HD 176377   &0.14		&...	&1	&0.19	&...	&1	&0.07	&0.06	&6	&0.14	&0.06	&2  &0.04	&0.03	&2	\\
\hline
\end{longtable}
}
\end{document}